\documentclass{IEEEtran}
\usepackage{amsmath, amssymb, amsthm}

\usepackage{graphicx, subcaption}
\usepackage[font=small]{caption}
\usepackage{cite}
\usepackage{xcolor}
 
\definecolor{hicol}{HTML}{000000}
\definecolor{hicol2}{HTML}{0277BD}
\usepackage{acronym}

    \acrodef{MPC}{multipath component}
    \acrodef{MIMO}{multiple-input and multiple-output}
    \acrodefplural{MPC}[MPCs]{multipath components}

    \acrodef{DoF}{Degree of Freedom}

    \acrodef{mMIMO}{massive MIMO}

    \acrodef{IRS}{intelligent reflecting surface}
    \acrodefplural{IRS}[IRSs]{intelligent reflecting surfaces}

    \acrodef{SV}{Saleh-Valenzuela}
    \acrodef{DK}{Dirichlet kernel}
    \acrodefplural{DK}[DKs]{Dirichlet kernels}

    \acrodef{ULA}{uniform linear array}
    \acrodefplural{ULA}[ULAs]{uniform linear arrays}
    \acrodef{UPA}{uniform Planer array}
    \acrodefplural{UPA}[UPAs]{uniform planer arrays}

\begin{document}
\bstctlcite{IEEEexample:BSTcontrol}

\title{IRS-Enabled Beam-Space Channel}
\author{Musab Alayasra, \textit{Student Member, IEEE}, 
H\"{u}seyin Arslan, \textit{Fellow, IEEE} 
\thanks{
© 2021 IEEE. Personal use of this material is permitted. Permission from IEEE must be obtained for all other uses, in any current or future media, including reprinting/republishing this material for advertising or promotional purposes, creating new collective works, for resale or redistribution to servers or lists, or reuse of any copyrighted component of this work in other works.

The work of H. Arslan was supported by the Scientific and Technological Research Council of Turkey (TUBITAK) under Grant 120C142

The authors are with the Department of Electrical and Electronics Engineering, Istanbul Medipol University, Istanbul, 34810, Turkey (e-mail: musab.alayasra@std.medipol.edu.tr).

H. Arslan is also with the Department of Electrical Engineering, University of South Florida, Tampa, FL, 33620, USA (e-mail: arslan@usf.edu).
}}

\maketitle

 \begin{abstract}
 The \ac{IRS} is emphasized as a controlled scattering cluster. To this end, scatterers and traveling paths of multipath components are classified to build a new channel model. Unlike the conventional modeling, where the channels between system units are modeled independently, the new model considers the channel as a whole and decomposes it based on the traveling paths. The model shows clearly how \ac{IRS}, in the beam-space context, converts the channel from a problem into a design element. After investigating \ac{IRS} as a scattering cluster, based on a proposed segmentation scheme, the beamforming problem is considered with a focus on first-order reflections. Passive beamforming at \ac{IRS} is shown to have two tiers; at the scatterer and antenna levels. A segment-activation scheme is proposed to maximize the received signal power, where the number of transmitting antenna elements to be used is given as a function of \ac{IRS} positioning and beamforming at the receiver. The results show that while using more transmitting antenna elements to get narrower beams is possible, using fewer elements can give better performance, especially for larger \ac{IRS} at close distances. The developed model also proves useful in addressing emerging issues in massive MIMO communication, namely, stationarity and spherical wavefronts.
 \end{abstract}


\section{Introduction}
\label{sec:intro}
\acresetall
\IEEEPARstart{W}ITH every new wireless communication generation, the demand for more resources and better services by emerging applications is accelerating. In response to their urging calls, new generations also come with novel solutions. One example is mmWave technology that proved to be a key player in enhancing spectral efficiency and accommodating demanding applications. Another prominent example is the recently proposed concept called \ac{IRS} \cite{8466374,8796365,Renzo_2020}. As a controlled reflecting object, \ac{IRS} can be thought of as a \textit{controlled scattering cluster} in the environment responsible for a group of \acp{MPC} reflected toward a receiver, which enables a degree of control over the wireless channel. This paradigm shift in wireless communication systems could open the door for promising solutions that were never possible before.

Many works on this topic have been recently reported to improve different aspects of the wireless communication system like information and secrecy rates \cite{zhang2020capacity,qiao2020secure,shen2019secrecy,cui2019secure}, power efficiency \cite{8811733,8741198,wang2020intelligent}, multiple accessing \cite{kammoun2020asymptotic,ding2020simple,hou2019reconfigurable}, and others \cite{9198898,9122596}. However, the channel model adopted by almost all works shows a common perspective that treats \ac{IRS} as a third communication unit along with transmitter and receiver without demonstrating clearly its role as a controlled object in the environment. As such, the fundamental antenna-domain channel model is adopted in describing the \ac{IRS}-enabled wireless channel as a cascade of two channels in addition to the main one, which is given as $\mathbf{G}\boldsymbol{\Phi}\mathbf{F}+\mathbf{H}_{\rm D}$, where $\mathbf{H}_{\rm D}$ is the channel between transmitter and receiver, $\mathbf{F}$ and $\mathbf{G}$ are, respectively, the channels between \ac{IRS} and each of transmitter and receiver, and $\boldsymbol{\Phi}$ is a diagonal matrix representing phase shifts introduced by \ac{IRS} elements. Besides, the estimation of these channels has an overhead issue due to the large number of antenna elements at \ac{IRS} \cite{9133142, zegrar2020general, zegrar2020reconfigurable}, and possibly at the transmitting and receiving units. The sparsity of mmWave channel can be exploited to reduce this overhead \cite{9127834,8879620}. However, as \ac{IRS} is envisioned to have large sizes covering walls and objects in the environment \cite{Renzo_2020}, the assumption of far-field operation is not practical and there could be a non-stationarity \cite{8424015} over the \ac{IRS} surface, where different segments of its elements experience different sets of scattering clusters with transmitter or receiver.

Millimeter waves have some features that make a difference in the context of \ac{IRS}. Their relatively shorter wavelengths result in higher path loss but at the same time allow packing more antenna elements into a compact size to generate narrower beams with higher beamforming gain \cite{6736750}. Directed beams and high path loss render mmWave channels sparser in general with less number of \acp{MPC} and lower-order reflections \cite{8424015,7109864}. For instance, at the $60$GHz band in indoor environments, the campaigns in \cite{peter2017measurement} show that the number of scattering clusters ranges from $2$ to $5$ clusters. At the same band, it is shown in \cite{maltsev2009experimental} that the power ratio of first to second-order reflection is similar to that of LoS path to first-order reflection with an approximate value of $13$dB. In addition to their definite advantage in terms of received power, focusing on first-order reflections in proposing \ac{IRS}-based solutions (i.e., reflections by \ac{IRS} only) reduces the overhead of channel estimation mentioned above. Since only LoS links with \ac{IRS} are utilized, small-scale fading information in $\mathbf{F}$ and $\mathbf{G}$ can be safely ignored. This fact is better understood and exploited if the channel is described geometrically.

At mmWave frequencies, thinking of the channel geometrically is essential to reduce implementation complexity \cite{zhang2020prospective}. Unlike the antenna-domain channel model, which treats the wireless channel as a black box whose inputs and outputs are antenna elements, geometric and beam-space channel models \cite{1033686,tse2005fundamentals,8207426} are more descriptive in telling what is inside the box. In the beam-space channel, inputs and outputs are beams instead of antenna elements, and they act as ports in the angular domain to which transmit power is allocated. Over this domain, the geometric model tells the distribution of those objects in the environment reflecting non-vanishing \acp{MPC} toward the receiver. \ac{IRS} is one of those objects, and those spatial dimensions, or beams, overlapping with its location are utilized in its reflections. On the other hand, fading over remaining dimensions needs to be estimated not for channels with \ac{IRS} but for the channel as a whole to assess other scattering clusters between transmitter and receiver. Modeling \ac{IRS} in the beam-space context is essential not only for a better understanding of its behavior but also to support non-traditional antenna hardware implementations \cite{zhang2020prospective}, where we think of the transmitting or receiving unit all as a single scattering system.

Although geometric modeling is considered for \ac{IRS}-enabled channels in modeling \cite{basar2020indoor}, channel estimation \cite{9127834,8879620}, and other works \cite{wang2020intelligent, qiao2020secure, wang2020joint}, they implicitly assume far-field operation and do not take into account the stationarity issue. Both near-field operation and stationarity need to be considered as \ac{IRS} can be large in size \cite{Renzo_2020}. However, if \ac{IRS} is considered as a scattering cluster with LoS links, stationarity is not a problem as the fading in $\mathbf{F}$ and $\mathbf{G}$ need not be estimated. But still, the far-field operation assumption is impractical, and the stationarity of \ac{IRS} itself at the transmitter and receiver needs to be considered. In \cite{wang2020joint}, \ac{IRS} is modeled differently but again as a scatterer in the far field, not a scattering cluster as done independently in this paper. Therefore, there is a need to thoroughly investigate \ac{IRS} as a scattering cluster and examine the characteristics of its reflected signal regardless of its field of operation.

To fill this gap, a new channel model is proposed based on geometric and beam-space channel models. The main idea is to segment antennas into smaller parts and recognize \ac{MPC} types in the channel. By segmentation, \ac{IRS} behavior as a scattering cluster is clearly understood. We also show the beam sub-spaces of the whole beam-space channel and how it can be turned from a problem into a design element in the system. Segmentation eliminates the spherical wavefront issue and enables grouping segments based on the visibility regions of the scattering clusters. One main feature of the model is that the splitting based on \ac{MPC} types enables the designer to select which parts of the channel to use for communication. For instance, if only the paths through \acp{IRS} are selected, there will be no need for channel estimation, but for a mobile receiver, beam training is required. Although we focus on first-order reflections by \ac{IRS}, other types of reflections are also modeled and discussed. Based on the developed model, beamforming is next considered. Passive beamforming at \ac{IRS} is shown to have two tiers; one at the scatterer level and the other at the antenna level with the scatterers as elements. A single-segment activation scheme at the transmitter side is proposed to maximize the received power. The number of antenna elements in this active segment is derived in a closed-form as a function of \ac{IRS} angular span seen by the transmitter and beamforming at the receiver.

Different approaches are followed in the literature for path loss calculation in \ac{IRS}-enabled wireless channels. The first one, as shown in \cite{ellingson2019path} and \cite{tang2020wireless}, is based on antenna theory, where \ac{IRS} is treated as an array of passive elements, each with a given radiation pattern. The overall path gain of reflected signals is found by superposing the path gains of those individual elements. The second approach is to find the electromagnetic field radiated by \ac{IRS} for a given incident wave, then based on its value, path loss can be calculated at any given point \cite{Renzo_2020,8936989}. In \cite{gradoni2021end} a circuit-based approach is adopted where \ac{IRS} elements are modeled as tunable impedances to control an equivalent channel that maps voltages at transmit and receive antenna elements. The antenna-theory-based approach is followed in this paper to build the channel model.
 
\subsection{Contribution}
The contribution of this work is summarized in the following points:
\begin{itemize}
 \item A classification of the scattering clusters and \acp{MPC} traveling paths in the channel is given. Also, a segmentation scheme is proposed to divide an antenna into smaller parts, called segments or scatterers. Based on the classified traveling paths, the channel is split into three sub-channels. Then, based on segmentation, the operation field issue is addressed by fitting the segments into the conventional channel models. The developed model shows clearly how the channel is controlled by IRS in the beam space. 
 \item By segmentation, IRS is presented and investigated as a controlled scattering cluster that fits very well in the geometric model. \ac{IRS} segments are equivalent to scatterers in a scattering cluster, and their gains are shown to be directly related to their phase profiles. This relation is described by what is referred to as the compensation gain.
 \item Based on the developed model and the segmentation scheme, the cascaded beamforming is next addressed to maximize the power delivered by \ac{IRS}. A segment-activation method is proposed to eliminate phase delays due to propagation so that the received \acp{MPC} add constructively.
\end{itemize}

The rest of the paper has the following sections. Section \ref{sec:mdl} presents the proposed beam-space model. In Section \ref{sec:IRS_CC}, IRS as a scattering cluster and its reflections are investigated. Other types of reflections are discussed in Section \ref{sec:othr_rfl}. Back to reflected signals by \ac{IRS}, a cascaded beamforming scheme is proposed in Section \ref{sec:near_bf} to maximize the received power. Also, simulations are given in this section to show the performance of the proposed solution. Finally, the paper is concluded in Section \ref{sec:cnc}.

\section{Channel Decomposition}
\label{sec:mdl}
In the developed model, there are three system units: transmitter $T_X$, receiver $R_X$, and \ac{IRS}, denoted by $L_X$. Their antennas are modeled as \acp{ULA}, where $T_X$ ($R_X$ or $L_X$) has $N_T$ ($N_R$ or $N_L$) antenna elements with antenna size $D_T$ ($D_R$ or $D_L$) and elements spacing $q_T$ ($q_R$ or $q_L$). Antenna elements are modeled as scatterers with a uniform radiation pattern in all directions except for $L_X$. \ac{IRS} is a metasurface, and metasurfaces can be either periodic or aperiodic \cite{yang2019surface,Renzo_2020}. In both cases, it is a lattice of unit cells, each with a given structure. Periodic metasurfaces have identical unit cells, and a prototype of such architecture is shown in \cite{tang2020wireless}. Depending on its structure, the gain of a unit cell might not be homogeneous at all angles, so the radiation pattern of $L_X$ elements is assumed to be of any shape and denoted by $B_{L,e}(\theta)$. As an example, it is proposed in \cite{ellingson2019path,tang2020wireless} to use $B_{L,e}(\theta) = a \cos^{b}(\theta)$ with $a$ and $b$ as related constants. In this work, we assume any radiation pattern for $L_X$ elements.

\subsection{Antenna Segmentation}
Antennas can have any size and may operate in the near fields of each other. Operation fields are defined based on the approximation of the spherical wavefronts as plane wavefronts \cite{8736783}. The spherical wavefront of a point source illuminating an antenna is approximated as a plane wave if it is farther than a given distance called far-field distance. This distance depends on the antenna size and the accepted maximum phase difference over its elements. For an antenna with a maximum dimension $D_T$, it should be more than or equal to $2D_T^2/\lambda$ for a maximum phase difference of $\pi/8$ \cite{7942128}.

\begin{figure}[t]
 \centering
\includegraphics[scale=0.45]{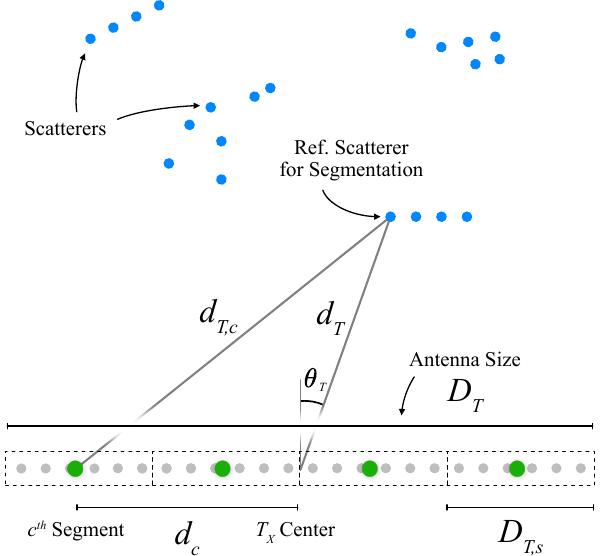}
\caption{\color{hicol}Antenna segmentation.}
\label{fig:seg0}
\end{figure}

To fit in the geometric channel model, antennas are divided into segments of elements. Similar to ordinary objects in the environment, these segments are also called scatterers. For the geometric model to be valid, scatterers with direct interaction must be in the far field of each other. Figure \ref{fig:seg0} shows an example of how to segment $T_X$ given its distance with other scatterers. Other antennas and environment objects form a constellation of scatterers seen by $T_X$. The nearest scatterer acts as a reference object used to divided $T_X$ into equal-size segments such that it will operate in their far fields. This requirement is met if
\begin{equation}
 \min_{c}(d_{T,c})\geq \frac{2D_{T,s}^2}{\lambda},
 \label{eq:seg_dis}
\end{equation}
where $d_{T,c}$ is the distance between the scatterer and the $c^{\text{th}}$ segment whose size is $D_{T,s}$. Therefore, a segmentation that results in $D_{T,s}$ satisfying (\ref{eq:seg_dis}) is valid. The use of the segmentation method is demonstrated in the next sub-section and later sections.

\subsection{Scatterers Splitting}
\begin{figure}[t]
 \centering
 \includegraphics[scale=0.4]{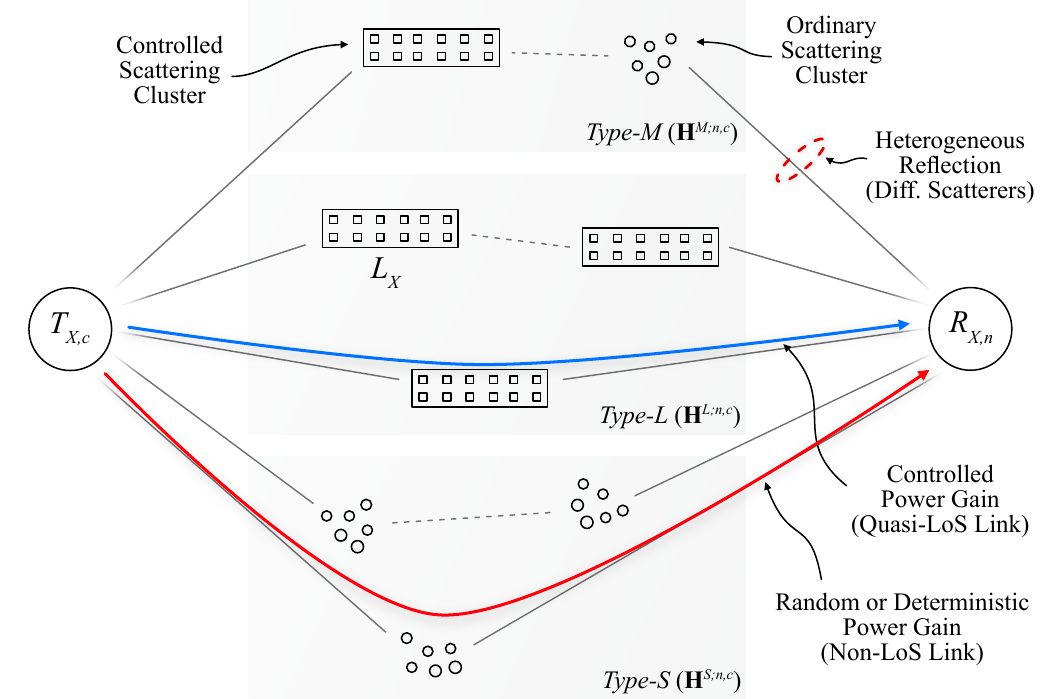}
 \caption{\textcolor{hicol}{Partial illustration of the geometric channel model with \ac{IRS} as a controlled scattering cluster.}}
 \label{fig:mdl_geo}
\end{figure}

The channel is assumed to be sparse with countable scatterers reflecting non-vanishing \acp{MPC} to the receiver. As shown in Fig. \ref{fig:mdl_geo}, scattering clusters in the environment are of two types: ordinary clusters (random objects in the environment) and controlled clusters (\acp{IRS} with their segments as scatterers; see Section \ref{sec:IRS_CC}). Hence, the path through which an \ac{MPC} travels is one of two types: homogeneous or heterogeneous. In the former, reflections of any order are due to one scatterer type; either controlled or ordinary scatterers. In the latter, reflections have a minimum order of two, and they are due to the two types of scatterers. As such, the channel matrix is decomposed into three matrices as
\begin{equation}
 \mathbf{H} = \mathbf{H}^S+\mathbf{H}^L+\mathbf{H}^M \in \mathbb{C}^{N_R\times N_T},
 \label{eq:mdl_dvd}
\end{equation}
where $\mathbf{H}^S$ corresponds to the homogeneous reflections by ordinary scatterers (type-$S$), $\mathbf{H}^L$ corresponds to homogeneous reflections by controlled scatterers (type-$L$), and $\mathbf{H}^M$ corresponds to heterogeneous reflections (type-$M$). If we assume that the objects where \acp{IRS} are mounted have no contribution to the received signal when \acp{IRS} are uninstalled, $\mathbf{H}^S$ is the ordinary channel of the \ac{MIMO} system.

The geometric model relies on plane waves to describe the channel. It treats transmitting, receiving, and reflecting objects as scatterers and describes them by the response vectors of their incident or emitted waves at specific angles. As mentioned before, the far-field operation for those scatterers with direct interaction is necessary. Therefore, we assume that $T_X$ and $R_X$ are divided into $S_T$ and $S_R$ segments, respectively, so that any scatterer is in their far field. Consequently, the matrices in (\ref{eq:mdl_dvd}) are given as block matrices where 
\begin{equation}
 \mathbf{H}^{y} = \begin{bmatrix} 
 \mathbf{H}^{y;1,1} & \dots & \mathbf{H}^{y;1,S_T}\\
 \vdots & \ddots & \\
 \mathbf{H}^{y;S_R,1} & & \mathbf{H}^{y;S_R,S_T} 
 \end{bmatrix}
 \label{eq:mdl_blks}
\end{equation}
with $y\in\{S,L,M\}$. The matrix $\mathbf{H}^{y;n,c}$, for $c=1,\dots,S_T$ and $n=1,\dots,S_R$, represents the type-$y$ channel between the $c^{\text{th}}$ transmitter segment, $T_{X,c}$, and the $n^{\text{th}}$ receiver segment, $R_{X,n}$. It is modeled using the extended \ac{SV} model as follows \cite{6717211}
\begin{equation}
 \begin{split}
 \mathbf{H}&^{y;n,c} = \\
  &\sum_{i,j} \beta^{y;n,c}_{j,i} \mathbf{a}_{R,S}(\theta^y_{R,n;i,j})(\mathbf{a}_{T,S}(\theta^y_{T,c;i,j}))^T \in \mathbb{C}^{N_{R,S}\times N_{T,S}}.
 \end{split}
 \label{eq:sctr_geo}
\end{equation} 
The $j^{th}$ scatterer in the $i^{th}$ cluster has a gain $\beta^{y;n,c}_{j,i}$, and the angles $\theta^y_{T,c;i,j}$ and $\theta^y_{R,n;i,j}$ describe its angular locations relative to $T_{X,c}$ and $R_{X,n}$ centers, respectively. Angular locations are with respect to the norm vector of the antenna aperture (i.e., using the broadside angle $-\pi/2\leq\theta\leq\pi/2$). The number of antenna elements in $T_{X,c}$ and $R_{X,n}$ are $N_{T,S}$ and $N_{R,S}$, respectively. The response vectors modeling the scatterers are given as
\begin{equation}
 \mathbf{a}_{x,S}(\theta^y_{x,c;i,j}) = \left[ e^{-j\gamma^y_{x,c;i,j}l} \right]_{l\in \mathcal{I}_{N_{x,S}}} \in \mathbb{C}^{N_{x,S}\times 1},
 \label{eq:bf_conv}
\end{equation}
where $x\in\{T,R\}$ and $c=1,\dots, S_x$. Given the wavelength $\lambda$ and wavenumber $k=2\pi/\lambda$, $\gamma^y_{x,c;i,j} = kq_x\sin(\theta^y_{x,c;i,j})$. Finally, the set $\mathcal{I}_z$ is defined as \cite{6484896}
\begin{equation}
 \mathcal{I}_z = \left\{ i-\frac{(z-1)}{2}; i=0,1,\dots,z-1 \right\}.
 \label{eq:int1}
\end{equation}

The response vector for a given scatterer over the whole transmit or receive antenna is a stack of the response vectors at its segments 
\begin{equation}
 \begin{split}
 \mathbf{a}_x(\theta^y_{x;i,j}) = \Big[ \mathbf{a}_{x,S}(\theta^y_{x,c;i,j}) \Big]_{c\in \{1,..,S_x\}} \in \mathbb{C}^{N_{x}\times 1},
 \end{split}
\end{equation}
where $\theta^y_{x;i,j}$ is the scatterer angular location relative to $x$. Note that the response vector of an antenna, as one segment, is decomposed into several response vectors of smaller segments. As shown later, in this manner, the geometric model will support the spherical wavefront as we divide it into several plane wavefronts at each segment. 

{\color{hicol}
Unlike the conventional channel model, splitting the channel based on the traveling paths shows clearly the controlled part of the channel. As seen in Fig. \ref{fig:mdl_geo}, $\mathbf{H}^{S}$ delivers purely fading components and no control is possible over this sub-channel. In contrast, fully controlled components travel over the paths in $\mathbf{H}^{L}$. Coming between these two sub-channels is $\mathbf{H}^{M}$ that carries fading components affected by beamforming at \ac{IRS}. The gain of a path in $\mathbf{H}^{M}$ is partially controlled due to the existence of ordinary scatterers; thus, it can be either deterministic or random depending on the nature of the ordinary scatterers. Finally, as it is possible to have \ac{IRS} common to different paths in $\mathbf{H}^{L}$ and $\mathbf{H}^{M}$, the design of $\mathbf{H}^{L}$ could affect the fading in $\mathbf{H}^{M}$, but $\mathbf{H}^{S}$ is always independent of $\mathbf{H}^{L}$ design.
}

It is important to mention that segmentation does not necessarily mean an increase in the number of unknowns to be estimated. For instance, in $\mathbf{H}^{L}$, the goal of segmentation is always to convert the spherical wavefront into multiple plane wavefronts for the same set of scatterers. Also, note that for $S_T = S_R = 1$, the channel matrix in (\ref{eq:mdl_blks}) is one-block, which is the conventional model widely used in the literature for \ac{MIMO} systems. Different segmentation sizes for different scattering clusters are possible. For example, ordinary scattering clusters between $T_X$ and $R_X$ might all operate in the far fields, but there is an \ac{IRS} in the near field. In such case, $S_T=S_R=1$ for $\mathbf{H}^S$, but multiple segments are required at $T_X$ and $R_X$ for $\mathbf{H}^L$.

The model in (\ref{eq:sctr_geo}) is for frequency-flat fading channels; however, the extension to frequency-selective channels is possible using OFDM-based precoding and combining \cite{7448873}. In this case, the same model applies but for individual subcarriers. Similar to analog beamforming at $T_X$ and $R_X$ sides, the same passive beamforming at \ac{IRS} applies to all subcarriers.

\subsection{Scatterers Gains}
The gains applied to \acp{MPC} reflected by ordinary scatterers, $\left\{\beta^S_{i,j}\right\}$, are widely modeled stochastically with a given distribution. However, other methodologies are followed in channel modeling to include \acp{MPC}, other than the LoS component, with deterministic power gains \cite{8424015}. For instance, the IEEE 802.11ay channel model \cite{maltsev2017channel} follows this quasi-deterministic approach to model the mmWave channel with different types of \acp{MPC}. One type is called D-ray, and it is found based on ray-tracing reconstruction for a given scenario environment. It is an \ac{MPC} reflected from a macro object with possibly added random \acp{MPC} to form a quasi-deterministic \ac{MPC}. Another type is called R-ray and it represents reflections from random objects. This type of \acp{MPC} is random and follows the complex Gaussian distribution. On the other hand, as shown in later sections, $\left\{\beta^L_{i,j}\right\}$ are controlled by means of passive beamforming at $L_X$ side, and they are always deterministic for known $T_X$ and $R_X$ locations.

Ordinary scattering clusters reflect signals with an uncontrolled gain that depends on their composition, but signals reflected by \ac{IRS} are optimized in the sense that reflected \acp{MPC} can be focused to maximize the received power. Also, \ac{IRS} gives a degree of control over those reflected \acp{MPC}. For example, it can direct the incident signal into different directions, which is a feature that is not possible by ordinary scattering clusters. However, optimizing \ac{IRS} reflections requires accurate-enough knowledge of its location relative to $T_X$ and $R_X$. For point-to-point communication with fixed $T_X$ and $R_X$ locations, as in backhaul communication, this might be a valid assumption. However, in the case of having mobile $R_X$, a (beam) training stage is required to compensate for phase differences between reflected \acp{MPC}. 

\subsection{Beam Subspaces}
\label{sec:mdl_subspcs}

\begin{figure}[t]
    \centering
    \includegraphics[scale=0.35]{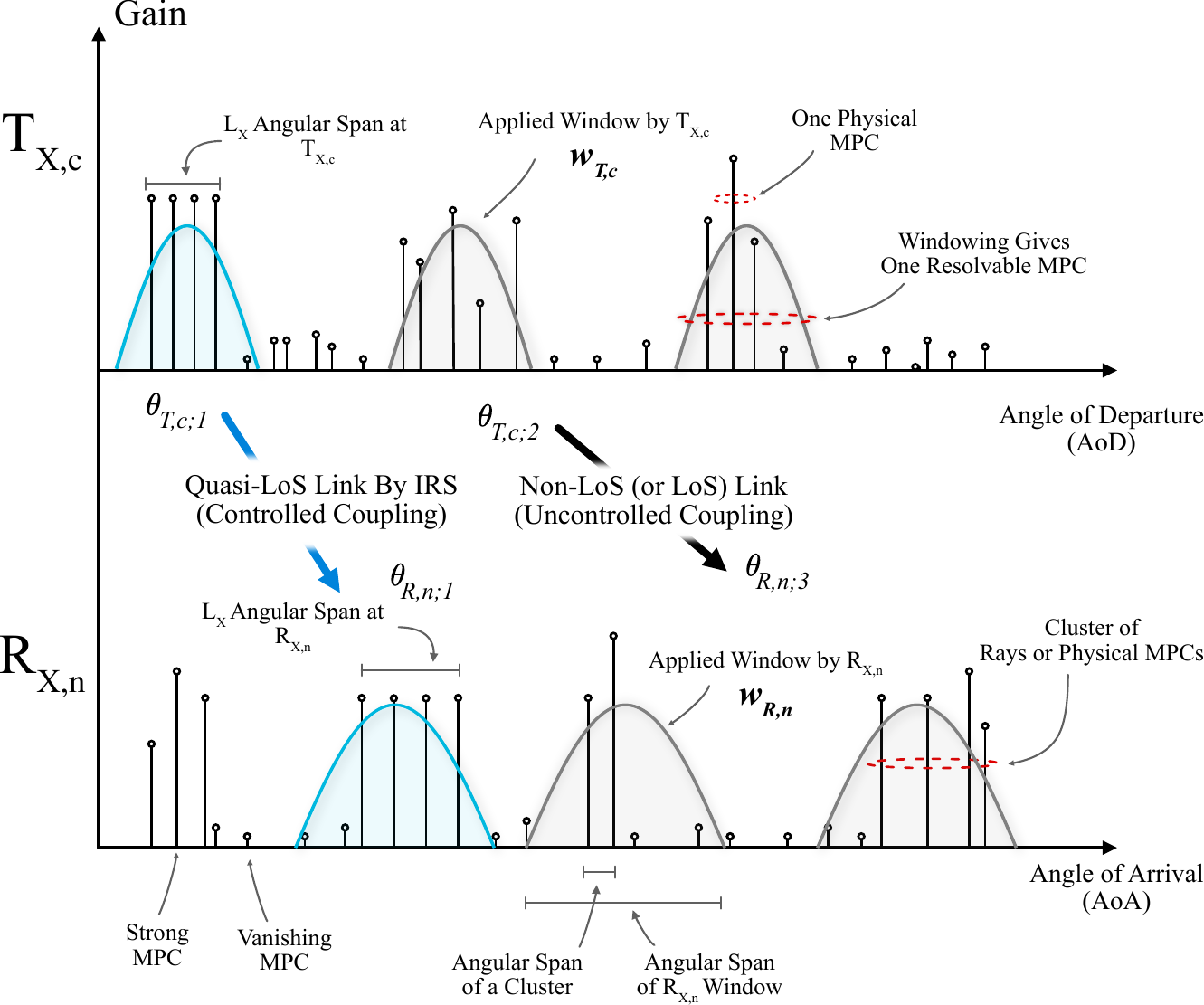}
    \caption{\color{hicol}Beam-space channel model with \ac{IRS} as a source of cluster of \acp{MPC}.}
    \label{fig:mdl_bm}
\end{figure}

Figure \ref{fig:mdl_bm} shows how geometric and beam-space channel models fit together.
Beamforming at $T_{X,c}$ or $R_{X,n}$ side is approximated by selection windows in the angular domain. 
The output of a window is one resolvable \ac{MPC}, which is a combination of unresolvable physical \acp{MPC} \cite{6848765}. (Here, we might also call a group of physical \acp{MPC} as one physical \ac{MPC}.) It is also shown in the figure how \ac{IRS} acts similar to other clusters by reflecting multiple \acp{MPC}, which are not random but controlled as shown later in Section \ref{sec:IRS_CC}.

Similar to response vectors, beamforming vectors are decomposed into smaller vectors for individual segments. Hence, for a given segment, the beam-space beamforming vector \cite{6484896} is 
\begin{equation}
    \mathbf{w}_{x,S}(\theta_{{x,c},S}) = \frac{1}{\sqrt{N_x}}\Big[ e^{j\gamma_{{x,c},S}l} \Big]_{l\in \mathcal{I}_{N_{x,c}}} \in \mathbb{C}^{N_{x,c}\times 1},
    \label{eq:bmsub_eq1}
\end{equation}
where $\gamma_{{x,c},S}=kq_x\sin(\theta_{{x,c},S})$ with $\theta_{{x,c},S}$ as the segment steering angle. This vector corresponds to a beam that acts as a selection window applied by the segment to select a specific angular span as shown in Fig. \ref{fig:mdl_bm}. Physical \acp{MPC} within a window go under different beamforming gains depending on the beam shape, and the gain at directions outside the window is small enough to be neglected.

The beamforming vector over the whole transmit or receive antenna is given as 
\begin{equation}
    \begin{split}
        \mathbf{w}_x = \Big[ \mathbf{w}_{x,S}(\theta_{x,c,S}) \Big]_{c\in \{1,..,S_x\}} \in \mathbb{C}^{N_{x}\times 1}.
    \end{split}
    \label{eq:mdl_bmsp1}
\end{equation}
For multiple RF chains, the analog beamforming matrix is given as
\begin{equation}
    \mathbf{W}_x = \Big[\mathbf{w}^1_{x}, \mathbf{w}^2_{x}, \dots, \mathbf{w}^{M_x}_{x}\Big] \in \mathbb{C}^{N_x\times M_x}
\end{equation}
where $\mathbf{w}^i_{x}$ is the $i^{th}$ beamforming vector as given in (\ref{eq:mdl_bmsp1}) and $M_x$ is the number of RF chains. 

The beam-space channel matrix is given as
\begin{equation}
    \begin{split}
        \mathbf{H}_B &= (\mathbf{W}_R)^T \mathbf{H} \mathbf{W}_T \\
        &= (\mathbf{W}_R)^T \mathbf{H}^L \mathbf{W}_T + \mathbf{H}^I_B \in \mathbb{C}^{M_R\times M_T}
    \end{split}
    \label{eq:mdl_bms}
\end{equation}
Unlike $\mathbf{H}$ in (\ref{eq:mdl_blks}), $\mathbf{H}_B$ is not a block matrix, but its internal structure, based on $\mathbf{H}$ and $\mathbf{W}_x$, is a composition of block matrices. In the conventional beam-space channel, the matrix elements represent coupling between one pair of beams. But in $\mathbf{H}_B$ here, they represent coupling between one pair of composite beams. One such beam has multiple beams stemming from antenna segments. Segments beams are unresolvable in $\mathbf{H}_B$, though they can be controlled at $T_X$ and $R_X$ sides. The control is possible by steering the beam (beamforming) or changing its size (segmentation), but once $T_X$ sends a beam into the channel, it is not controlled unless it is sent through a homogeneous path of controlled scattering clusters. \textit{Therefore, by $L_X$ in the beam space, complete control of the information-carrying signal from source to destination is possible with almost no randomness.} In other words, $L_X$ converts the channel from a problem to a design element in the wireless communication system. This is shown clearly by the second line of (\ref{eq:mdl_bms}), where $\mathbf{H}^L$ is singled out as a controlled part of the geometric channel while other parts are left in  $\mathbf{H}_B^I$ as an interference source in the beam-space channel. A designer might consider beamforming design and \acp{IRS} distribution in the environment to suppress $\mathbf{H}_B^I$ while maintaining the \ac{IRS}-enabled channel part for communication. Such an approach can eliminate the need for channel estimation, but it might give rise to the need for beam training, mainly at $L_X$, to maximum reflected signal power. In Section \ref{sec:near_bf}, the two-tier nature of passive beamforming at $L_X$ explains this clearly, but first, we show how $L_X$ acts as a controlled scattering cluster in the next section.

\section{IRS as Scattering Cluster}
\label{sec:IRS_CC}
\color{hicol}
\begin{figure*}[t]
    \centering
    \includegraphics[width=0.95\textwidth]{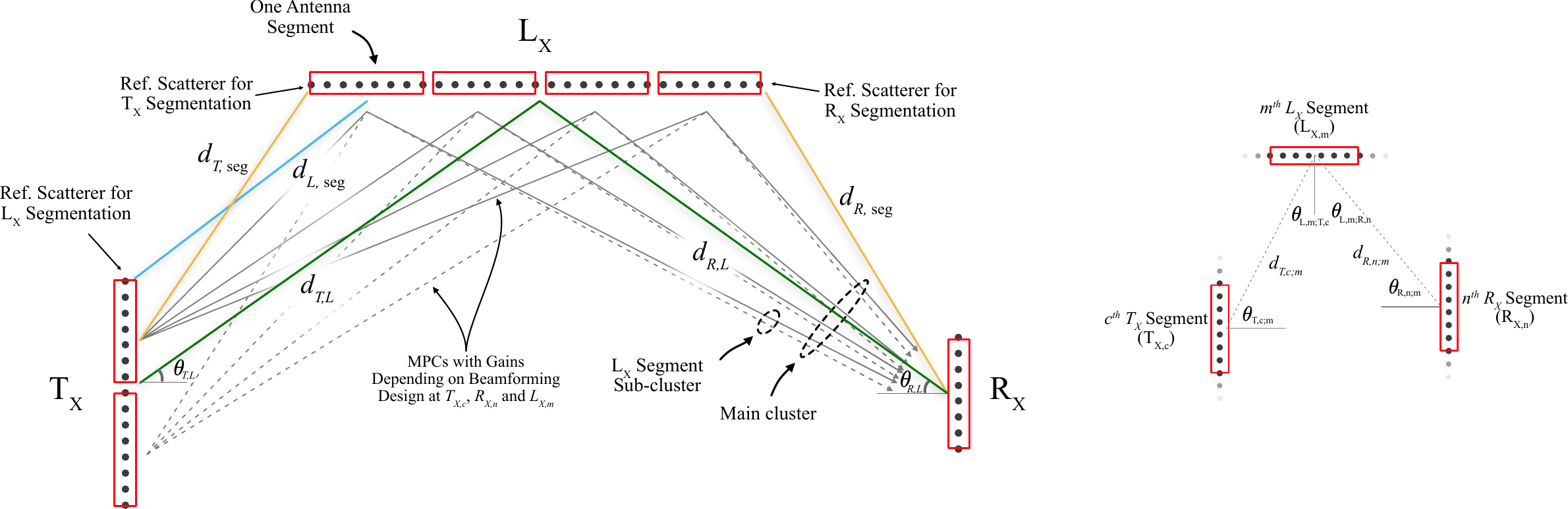}
    \caption{Antennas segments with their corresponding \acp{MPC}.}
    \label{fig:segs}
\end{figure*}

In this section, $\mathbf{H}^L$ in (\ref{eq:mdl_dvd}) is further investigated, where we study the gains of \acp{MPC} reflected by $L_X$ with a focus on first-order reflections. We assume a single \ac{IRS} in the channel, and the locations of $T_X$ and $R_X$ relative to $L_X$ are known.

We start with segmenting the system units. As shown in Fig. \ref{fig:segs}, $T_X$ interacts with $L_X$ only, so it is segmented based on its nearest scatterer at $L_X$, where we consider each $L_X$ element as a scatterer. This reference scatterer happened to be the edge element in this example and based on (\ref{eq:seg_dis}), we find the maximum $T_X$ segment size using its distance with the closest segment, which is $d_{T,\text{seg}}$. The same applies to $R_X$. On the other hand, $L_X$ interacts with both $T_X$ and $R_X$, so it is segmented based on the distances with their elements. In this example, it happened to have the reference scatterer at $T_X$. With the obtained segments as scatterers, $L_X$ is presented next as a controlled scattering cluster.

\color{black}
\subsection{IRS Multipath Components}
\label{sec:irs_mlt_cmps}
Consider the channel $\mathbf{H}^{L;n,c}$ between $T_{X,c}$ and $R_{X,n}$, and assume $S_L$ segments at $L_X$. Let the distance between the $m^{th}$ $L_X$ segment, $L_{X,m}$, and $T_{X,c}$ (or $R_{X,n}$) be $d_{T,c;m}$ (or $d_{R,n;m}$). The channel response from $T_X$ to $R_X$ is given as 
\begin{equation}
    \begin{split}
        \mathbf{H}^{L;n,c} = \sum_{m=1}^{S_{L}} \rho_{m;c,n} \mathbf{a}_{R,S}(\theta_{R,n;m}) (\mathbf{a}_{L,S}(\theta_{L,m;R,n}))^T \times&\\
        \boldsymbol{\Phi}_m \mathbf{a}_{L,S}(\theta_{L,m;T,c}) (\mathbf{a}_{T,S}(\theta_{T,c;m}))^T \in \mathbb{C}^{N_{R,S}\times N_{T,S}},&
    \end{split}
    \label{eq:sec2:clst}
\end{equation} 
where $\mathbf{a}_{T,S}(\theta)$, $\mathbf{a}_{L,S}(\theta)$ and $\mathbf{a}_{R,S}(\theta)$ are the channel response vectors at $T_{X,c}$, $L_{X,m}$ and $R_{X,n}$, respectively, as defined in (\ref{eq:bf_conv}). The angles $\theta_{T,c;m}$ and $\theta_{R,n;m}$ are the locations of $L_{X,m}$ with respect to the center of $T_{X,c}$ and $R_{X,n}$, respectively. On the other hand, $\theta_{L,m;T,c}$ and $\theta_{L,m;R,n}$ are the locations of $T_{X,c}$ and $R_{X,n}$, respectively, relative to the center of $L_{X,m}$ (See Fig. \ref{fig:segs}). The propagation coefficient given by 
\begin{equation}
    \rho_{m;c,n}=\frac{e^{jk(d_{T,c;m}+d_{R,n;m})}}{b_{\rm att}^2(d_{T,c;m}d_{R,n;m})^{a_{\rm att}/2}}
\end{equation}
describes path loss and phase delay due to propagation, where $a_{\rm att}$ is the path-loss exponent and $b_{\rm att}$ is a model-dependent constant. For free-space propagation, we have $a_{\rm att} = 2$ and $b_{\rm att} = 2k$. Finally, the matrix 
\begin{equation}
    \boldsymbol{\Phi}_m = diag(e^{j\phi_{m,1}},\dots,e^{j\phi_{m,N_{L,S}}}) \in \mathbb{C}^{N_{L,S}\times N_{L,S}}    
    \label{eq:seg_phss}
\end{equation}
represents the phases applied by $L_{X,m}$, where $\phi_{m,i}$ is the phase introduced by its $i^{th}$ element.

The term 
\begin{equation}
    \alpha_{m;c,n} = (\mathbf{a}_{L,S}(\theta_{L,m;R,n}))^T \boldsymbol{\Phi}_m \mathbf{a}_{L,S}(\theta_{L,m;T,c})    
    \label{eq:Lx_com_gain}
\end{equation}
is a scalar and called the \textit{compensation gain} of $L_{X,m}$ at $R_{X,n}$ for $T_{X,c}$ signal reflection, which depends on the segment passive beamforming design as shown in Section \ref{sec:near_bf}. If we define $B_{m;c,n} = B_{L,e}(\theta_{L,m;T,c})B_{L,e}(\theta_{L,m;R,n})$, then by writing the \ac{MPC} gain as
\begin{equation}
    \beta^{L;n,c}_{m} = B_{m;c,n}\rho_{m;c,n} \alpha_{m;c,n},
    \label{eq:mpc_gain}
\end{equation}
the channel in (\ref{eq:sec2:clst}) can be written as
\begin{equation}
    \mathbf{H}^{L;n,c} = \sum_{m=1}^{S_{L}} \beta^{L;n,c}_{m} \mathbf{a}_{R,S}(\theta_{R,n;m}) (\mathbf{a}_{T,S}(\theta_{T,c;m}))^T,
    \label{eq:irs_mpcs}
\end{equation}
which matches with the definition given in (\ref{eq:sctr_geo}) for a single scattering cluster, and it is the case when $T_X$ and $R_X$ have a small enough number of antenna elements so that any $L_X$ element operates in their far field. It is mainly the case of conventional \ac{MIMO} systems. Based on this result, the following comments are given:
\begin{itemize}
    \item IRS acts similar to ordinary scattering clusters; it is a collection of scatterers called segments. One \ac{MPC} stems from each segment; however, unlike those stem from ordinary scatterers, its gain is known and controlled.
    \item Equation (\ref{eq:irs_mpcs}) emphasizes IRS as a means to control the channel and signal propagation through it. As part of the channel, the gains of its reflected \acp{MPC} depend on its beamforming design. This is in contrast to beamforming at $T_X$ or $R_X$, which does not alter the channel but gives a response to it. 
    \item The minimum number of segments, or controlled scatterers, is governed by the largest permissible segment size found by (\ref{eq:seg_dis}). More segments might be assumed, but their number does not exceed $N_L$, where one segment is an individual element.
\end{itemize}

\color{hicol}
Additional segments at $T_X$ means additional \acp{MPC} reflected by $L_{X}$ with a total of $S_TS_L$ \acp{MPC}, as shown in the example in Fig. \ref{fig:segs}. Every $S_T$ \acp{MPC} reflected by $L_{X,m}$ are controlled as one group or sub-cluster and can not be distinguished by $L_{X,m}$. At $R_X$, having additional segments means receiving the $S_TS_L$ \acp{MPC} by $S_R$ segments with different gains. In general, the $S_TS_LS_R$ scatterers in $\mathbf{H}^{L}$ as given by (\ref{eq:mdl_blks}) and (\ref{eq:sctr_geo}) are dependent and correspond to the same $S_L$ scatterers but at different angles. \textit{As a result, only the change in $L_X$ segments changes the channel.} The scatterer drifting in angular domains seen by adjacent segments results in different gains of its reflected signal toward those segments, and differences are expressed by the compensation gains. The following sub-section reveals what is the \ac{MPC} reflected by \ac{IRS} in the beam-space context. But as a short answer, it is a weighted beam(s); more precisely, it is a cascade of plane waves.

\color{black}
\subsection{Geometric Interpretation}
\label{sec:geo}

\textcolor{hicol}{In the conventional beam-space channel, the beamforming design given in (\ref{eq:bmsub_eq1}) is adopted, but $T_X$ is one segment. For a steering angle $\theta_{T,S}$, the normalized field signal of $T_X$ at an angle $\theta_T$ is given as}
\begin{equation}
    \color{hicol}
    \begin{split}
        B_T(\theta_T;\theta_{T,S}) =& g_0(\mathbf{a}_T(\theta_T))^T \mathbf{w}_T \\ 
        =& 
        \frac{g_0}{\sqrt{N_T}}\sum_{l\in \mathcal{I}_{N_T}}e^{j(\gamma_{T,S}-\gamma_T)l}\\
        =& \frac{g_0}{\sqrt{N_T}}f(\gamma_T-\gamma_{T,S};N_T),
        \label{eq:segs1_tx_1}
    \end{split}
\end{equation}
where $g_0 = e^{jkd_T}/(d_T)^{a_{\text{att}}/2}$, and for the rest of the paper, we define $\gamma_{X,y} = kq_X\sin(\theta_{X,y})$ for $X\in\{T,R,L\}$ and any $y$. The function $f(\theta;N)$ is known as the Dirichlet kernel \cite{gradshteyn2014table} and it is an even function defined as
\begin{equation}
    f(\theta;N) = \sum_{l\in \mathcal{I}_N}e^{-j\theta l} = \frac{\sin(N\theta/2)}{\sin(\theta/2)},
    \label{eq:dk_def}
\end{equation}
We say that $f(\theta-\theta_o;N)$ has a size $N$ and is centered at $\theta_o$, which means its main lobe with a width $4\pi/N$ is directed toward $\theta_o$. 

\begin{figure}[t]
    \centering
\subfloat[]{\includegraphics[scale=0.4]{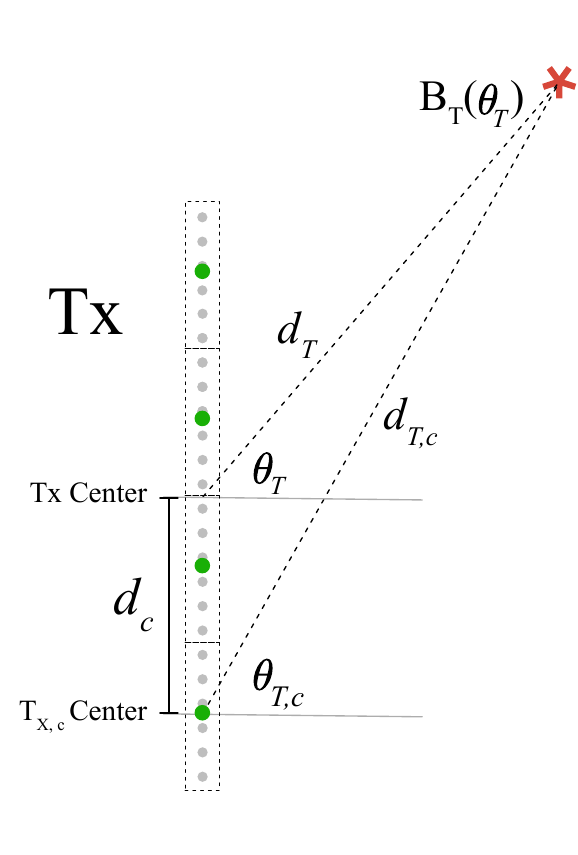}}
~~~
\subfloat[]{\includegraphics[scale=0.3]{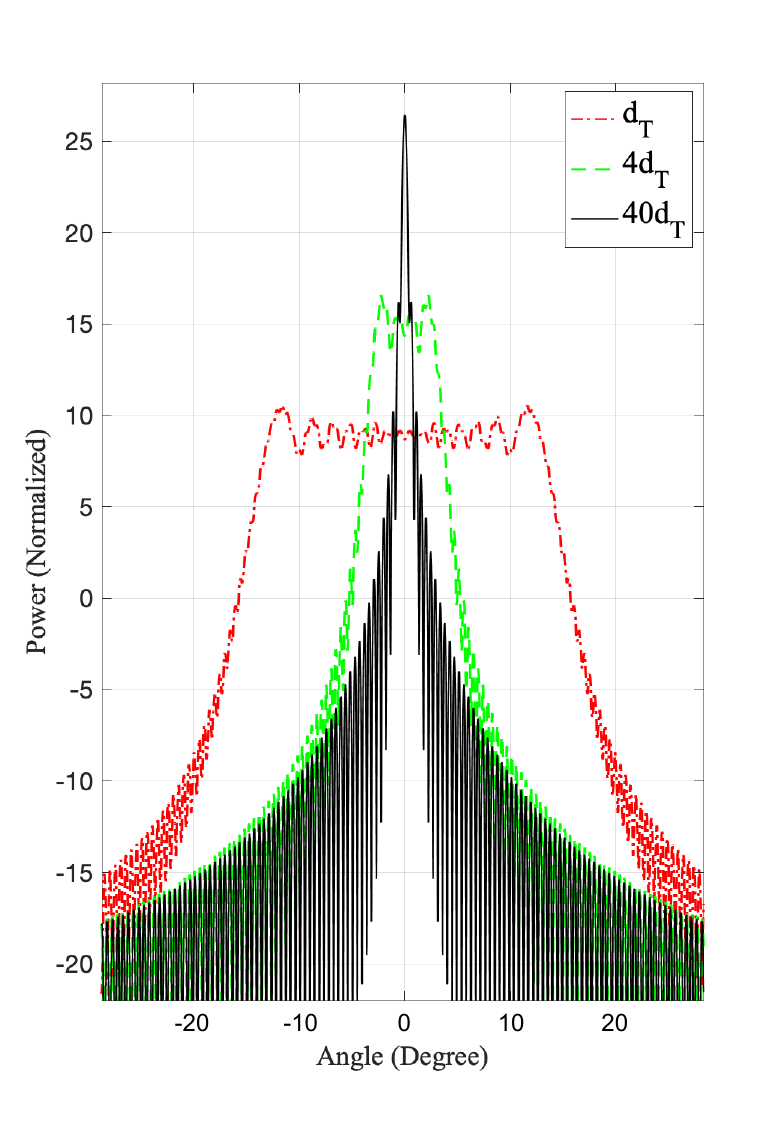}}
\caption{The beam convergence over distance.}
\label{fig:segs_flds}
\end{figure}

\textcolor{hicol}{In the near field, the field signal is calculated using the response vectors of individual segments. As shown in Fig. \ref{fig:segs_flds}(a), a point located at distance $d_T$ and angle $\theta_T$ with respect to $T_X$ center is located at $\theta_{T,c}$ and distance $d_{T,c}$ with respect to $T_{X,c}$ center. For the same conventional beamforming design as in (\ref{eq:segs1_tx_1}), we note that each segment beamforming vector has the same steering angle value ($\theta_{T,c,S}=\theta_{T,S}\forall c$), but there is a common phase applied to all its elements. Its value is $c\gamma_{T,S}N_{T,S}$ for $T_{X,c}$. The normalized field signal now might be expressed as}
\begin{equation}
    \color{hicol}
    \begin{split}    
        B_T&(\theta_T;\theta_{T,S})=\\
        &\sum_{c\in \mathcal{I}_{S_T}} (\underbrace{g'_{c}\mathbf{a}_{T,S}(\theta_{T,c})}_\text{Channel Response})^T\underbrace{\mathbf{w}_{T,S}(\theta_{T,S})e^{-jc\gamma_{T,S}N_{T,S}}}_\text{Beamforming Vector}
    \end{split}
\end{equation}
\textcolor{hicol}{where $g'_{c}= e^{jkd_{T,c}}/(d_{T,c})^{a_{\text{att}}/2}$. Similar to beamforming vectors, we notice that the channel response has the same two-tier feature, where each segment has a common phase delay to all its elements due to propagation. 
}

After simple manipulations, we get
\begin{equation}
    \color{hicol}
    \begin{split}
        B_T(\theta_T,\theta_{T,S})=
        \sum_{c\in\mathcal{I}_{S_T}} g_c
        f(\gamma_{T,c}-\gamma_{T,S}; N_{T,S}),
        \label{eq:segs_tx_2}
    \end{split} 
\end{equation}
where $g_c = g'_c e^{-jc\gamma_{T,S}N_{T,S}}/\sqrt{N_{T}}$. The above equation suggests treating each antenna segment as an independent source, which is the approach followed in this work. The field signal at a given point in the space is a superposition of weighted kernels of size $N_{T,S}$. 
It can be easily found that (\ref{eq:segs_tx_2}) converges to (\ref{eq:segs1_tx_1}) as $d_T\rightarrow \infty$. Figure \ref{fig:segs_flds}(b) shows how the beam evolves over distance as the kernels converge to one beam that is independent of distance.

\begin{figure}[t]
    \centering
\includegraphics[scale=0.6]{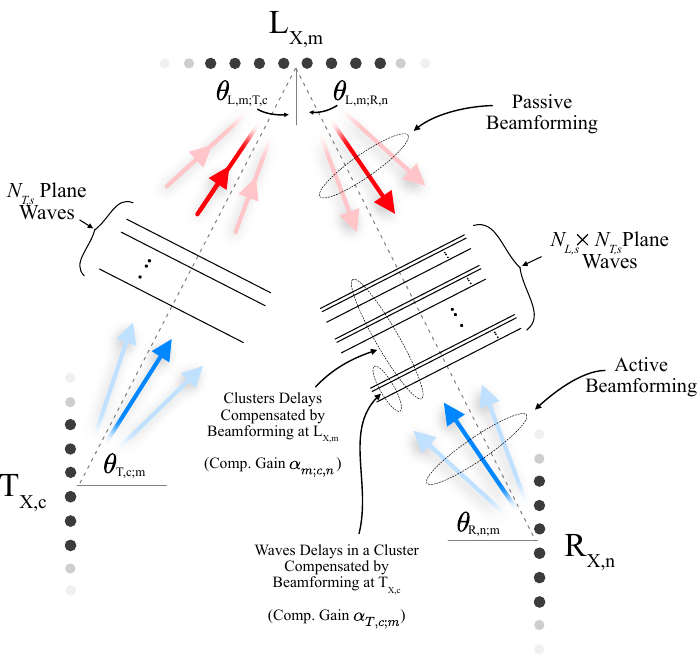}
\caption{Type-$L$ Reflections.}
\label{fig:actv_psv}
\end{figure}

Assume having only one $T_{X,c}$ in (\ref{eq:segs_tx_2}) and ignore its phase delay and path loss for now. Based on the kernel definition in (\ref{eq:dk_def}), $T_{X,c}$ acts as a source of $N_{T,S}$ plane waves arriving $L_{X,m}$ at the same angle $\theta_{L,m;T,c}$ as shown in Fig. \ref{fig:actv_psv}. 
Each plane wave can be thought of as a virtual source (i.e. single RF chain and its phase-shifting circuit) connected to $L_{X,m}$ elements and set to a steering angle $-\theta_{L,m;T,c}$, given that $\boldsymbol{\Phi}_m = \mathbf{I}_m$, where $\mathbf{I}_m$ is an identity matrix of size $m$. However, the steering angle can be controlled by $\boldsymbol{\Phi}_m$. For instance, if it is desired to direct the reflected beam toward $\theta_{L,m_o}$, the phases are set such that $\phi_{m,i} = \zeta_mi$ where $\zeta_m = \gamma_{L,m;T,c}+\gamma_{L,m_o}$. Thus, the reflected field signal, while ignoring the radiation pattern, is given as 
\begin{equation}
    \begin{split}
        &B_{L,m;c}(\theta_{L,m}) = \\
        &\hphantom{B}\sum_{l\in \mathcal{I}_{N_{T,S}}} \frac{e^{j(\gamma_{T,S}-\gamma_{T,c;m})l}}{\sqrt{N_{T}}} f(\gamma_{L,m}+\gamma_{L,m;T,c}-\zeta_{m}; N_{L,S}) \\
        &\hphantom{B_{L,m;c}(\theta_{L,m})}= \alpha_{T,c;m} f(\gamma_{L,m}+\gamma_{L,m;T,c}-\zeta_{m}; N_{L,S})
        \label{eq:geo_ref}
    \end{split}
\end{equation}
where $\theta_{L,m}$ is the angle with respect to $L_{X,m}$ center and $\alpha_{T,c;m} = f(\gamma_{T,S}-\gamma_{T,c;m};N_{T,S})/\sqrt{N_{T}}$ is the compensation gain of $T_{X,c}$ at $L_{X,m}$, which is defined in a similar manner to that for $L_{X,m}$ in (\ref{eq:Lx_com_gain}). Note that in the conventional beam-space channel, each system unit has one segment, so we have one compensation gain that can be maximized to have $\alpha_{T,c;m}^2 = N_{T,S}=N_T$. However, in near-field operation, $L_X$ might have multiple segments, and only one $L_{X,m}$ can have a maximum compensation gain.

Based on (\ref{eq:geo_ref}), we note that the reflected signal by $L_{X,m}$ is also a kernel. Therefore, the kernel at $T_{X,c}$ is equivalent to another one at $L_{X,m}$ but with different size, allowing for more design freedom (by having $T_X$ spanning angular domains of two different locations at the same time!)

The received signal by $R_{X,n}$ is a stack of clustered plane waves as seen in Fig.  \ref{fig:actv_psv}. To obtain maximum gain at $R_{X,n}$, it is necessary, but not sufficient, to compensate their phase differences. Each cluster has replicas of a plane wave received by $L_{X,m}$ from one element at $T_{X,c}$. Therefore, their phase differences are compensated by active beamforming at $T_X$, while phase differences between the clusters themselves is compensated by the passive beamforming at $L_{X,m}$.

In case of having multiple segments at $T_X$, each of them corresponds to an additional kernel reflected by $L_{X,m}$ at its own angle $-\theta_{L,m;T,c}$. Based on (\ref{eq:segs_tx_2}) and (\ref{eq:geo_ref}), the reflected signal by $L_{X,m}$ due to signals received from multiple segments at $T_X$ with the same steering angle is given as
\begin{equation}
    \color{hicol}
    \begin{split}
        B_{L,m}(\theta_{L,m}) =
        \sum_{c\in\mathcal{I}_{S_T}}\sqrt{N_{T,S}} g_c
        B_{L,m;c}(\theta_{L,m}).
    \end{split}
    \label{eq:geo_mult_tx_one_lx}
\end{equation}
Steering at $L_{X,m}$ is applied to all its kernels as one group, and it is not possible to steer them individually. The total reflected signal by $L_X$ can be found in a similar manner to the derivation of (\ref{eq:segs_tx_2}), where segments signals as defined in (\ref{eq:geo_mult_tx_one_lx}) can be controlled independently.

As a summary, in conventional beam-space modeling, $T_X$ with one RF chain is designed to give one beam in a given direction. On the other hand, $L_X$ gives $S_TS_L$ weighted kernels with a size that depends on the segment size. Every segment at $L_X$ reflects $S_T$ kernels and the difference between their angular directions depends on $T_X$ segments angular locations with respect to $L_{X,m}$. Thus, $T_X$ segments are equivalent to $S_T$ virtual sources connected to $L_{X,m}$ each with its own default steering angle $-\theta_{L,m;T,c}$. From a stationarity perspective, \ac{IRS} as a scattering cluster is seen at slightly different angles by $T_X$ segments. If $L_X$ is in the far field of $T_X$ and $R_X$ (i.e., each has one segment), the reflected signal by $L_{X,m}$ will be a single beam. Moreover, if $T_X$ and $R_X$ are in the far field of $L_X$, it acts as one scatterer with a single DK, which the case adopted in works like \cite{wang2020joint,wang2020intelligent}. Finally, the number of reflected kernels by $L_X$ increases by choosing smaller segment size; however, each new one will have a wider span and less reflected power.

\section{Other Reflections}
\color{hicol}
\label{sec:othr_rfl}
In previous sections, we focused on $\mathbf{H}^L$ with first-order reflections. In this section, higher-order reflections in $\mathbf{H}^L$ and other types of paths in (\ref{eq:mdl_dvd}) are discussed by following the same segmentation approach.

\subsection{Type-$S$ Reflections}
Two problems that arise for large antennas are non-stationarity and spherical wavefront\cite{8424015}. Implicitly, these problems were addressed for reflections by $L_X$. Ordinary scattering clusters, on the other hand, are not (or less) controlled and might not exist across the entire antenna \cite{6810277}. Consequently, the blocks in $\mathbf{H}^S$ might not have the same set of scatterers. Based on the concept of visibility regions \cite{6810277}, a cluster is common to a group of antenna elements if they fall in its visibility region. Therefore, ordinary scattering clusters exist in $\{\mathbf{H}^{S,n,c}\}$ based on their visibility regions. For the statistical modeling of visibility regions, \cite{6810277,8424015,8290972}, and references therein might be consulted. 

For a scatterer common to a large portion of the antenna, the spherical wavefront issue arises, and in the literature, it is addressed by modifying the response vectors to include phase delays due to the wave shape as shown in \cite{8290972}. In \cite{8290972}, also an approximation for the wave by a parabolic wavefront is given. Unlike spherical and parabolic wavefronts, segmentation maintains the connection between the geometric and beam-space models with the new interpretation discussed earlier by replacing the spherical wavefront with multiple plane wavefronts. Segmentation does not require the exact distance with a scatterer. It is possible to segment based on a distance below which no scattering clusters exist, but this distance should be scenario-dependent as it can be different for different environments.

\subsection{Higher-Order Reflections}
The extension to higher-order reflections in $\mathbf{H}^L$ is as follows. As the extension to other orders follows similarly, we assume a second-order reflection by \acp{IRS}. The first \ac{IRS}, $L_{X1}$, has interaction with $T_X$ and the second \ac{IRS}, $L_{X2}$, while $L_{X2}$ interacts with $R_X$ and not $T_X$. Therefore, $L_{X1}$ is segmented into $S_{L1}$ segments based on its distances with $T_X$ and $L_{X2}$, and $L_{X2}$ is segmented into $S_{L2}$ segments based on its distances with $L_{X1}$ and $R_X$. The number of segments at both \acp{IRS} is chosen as $S_L = \max (S_{L1},S_{L2})$. The same channel model in (\ref{eq:sctr_geo}) applies to the path with two \acp{IRS} as if they are one virtual \ac{IRS} with $S_L$ segments. The angles $\left\{\theta^L_{T,c;i,j}\right\}$ represent the locations of $L_{X1}$ segments with respect to $T_{X,c}$, and $\left\{\theta^L_{R,n;i,j}\right\}$ are the locations of $L_{X2}$ segments with respect to $R_{X,n}$. As each scatterer at $L_{X2}$ reflects signals coming from all scatterers at $L_{X1}$, the gains are given as
\begin{equation}
    \beta^{L;n,c}_{i,j} = \sum_{k=1}^{S_L} \beta^{L;n,c}_{i,j,k}
\end{equation}
where $\beta^{L;n,c}_{i,j,k}$ is the gain of the path from $T_{X,c}$ to $R_{X,n}$ through the $k^{\text{th}}$ and $j^{\text{th}}$ segments at $L_{X1}$ and $L_{X2}$, respectively, which can be easily found by following the approach in Section \ref{sec:irs_mlt_cmps}.

The heterogeneous paths in $\mathbf{H}^{M}$ give at least second-order reflections. Assume a second-order reflection, for simplicity, where there are two cascaded scattering clusters, one is ordinary and the other is controlled. As discussed above, it is possible to have a group of $L_X$ segments in the visibility region of the ordinary cluster. Only this subset of $L_X$ segments will contribute to $\mathbf{H}^{M}$. Steering vectors are defined similar to second-order reflections by \acp{IRS}; however, the gain of a path through the $j^{\text{th}}$ segment at $L_X$ and $i^{\text{th}}$ ordinary scatterer will be $\beta^{M,n,c}_{i,j} = \beta'_{i,j}\beta^{L,n,c}_{i,j}$, where $\beta'_{i,j}$ is the ordinary scatterer gain and $\beta^{n,c}_{L,i,j}$ is the segment gain as defined in (\ref{eq:mpc_gain}) but with respect to the scatterer and not $T_{X,c}$.

\color{black}
\section{Cascaded Beamforming}
\label{sec:near_bf}
Near-field beamforming to maximize the received signal power at $R_X$ through $L_X$ quasi-LoS link is addressed in this section. We assume analog beamforming at both $T_X$ and $R_X$ sides and that $R_X$ has only one segment. Also, we focus back on first-order reflection by \ac{IRS}.

\subsection{Problem Formulation}
\textcolor{hicol}{By considering one \ac{IRS} only in the environment, the channel between the $T_{X,c}$ and $R_X$ is given based on (\ref{eq:irs_mpcs}) as}
\begin{equation}
    \color{hicol}
    \mathbf{H}^{L;c} = \sum_{m=1}^{S_L} \beta^{L;c}_{m} \mathbf{a}_{R}(\theta_{R;m})(\mathbf{a}_{T,S}(\theta_{T,c;m}))^T.
\end{equation}
\textcolor{hicol}{As we focus on received power maximization, assume a unity transmitted signal with no information. Then, the received signal after combining is given based on (\ref{eq:mdl_bms}) as}
\begin{equation}
    \color{hicol}
    \begin{split}
        r =& (\mathbf{w}_{R})^T\sum_{c=1}^{S_T} \mathbf{H}^{L;c} \mathbf{w}_{T,c},
        \\
        =&\sum_{m=1}^{S_L} \sum_{c=1}^{S_T} \beta^{L;c}_{m}(\mathbf{w}_{R})^T \mathbf{a}_{R}(\theta_{R;m})
        (\mathbf{a}_{T,S}(\theta_{T,c;m}))^T \mathbf{w}_{T,c},
    \end{split}
\end{equation}
\textcolor{hicol}{where $\mathbf{w}_{T,c}= \left[ w_{T,c,i} \right]_{i\in\mathcal{I}_{N_{T,S}}}$ and $\mathbf{w}_{R}= \left[ w_{R,i} \right]_{i\in\mathcal{I}_{N_{R}}}$ are the beamforming vectors of $T_{X,c}$ and $R_{X}$, respectively. Therefore, the beamforming problem is given as}
\begin{equation}
    \color{hicol}
    \begin{split}
        \boldsymbol{\Gamma} &= \arg \max_{\boldsymbol{\Gamma}} |r|^2\\
        &\text{s.t. } |w_{R,i}|,|w_{T,c,i}|=1; \forall i
        \label{eq:opt1}
    \end{split}
\end{equation}
\textcolor{hicol}{where $\boldsymbol{\Gamma} = \left\{ \mathbf{w}_{T,c},\mathbf{w}_{R}, \boldsymbol{\Phi}_m \right\}_{m,c}$ is the beamforming parameters vector. }

We have two beamforming levels. Tier-1 beamforming is for the segments themselves and it is based on the design in (\ref{eq:bf_conv}). The second one is for the antenna with segments as its elements, and it is represented by a common phase applied to all segment elements. Therefore, we have
\begin{align}
    \color{hicol}
    \mathbf{w}_{T,c} &\color{hicol}= \mathbf{w}'_{T,S}(\theta_{T,c,S}) e^{j\phi_{T,c}}\\
    \color{hicol}
    \mathbf{w}_{R} &\color{hicol}= \mathbf{w}'_{R}(\theta_{R,S}) e^{j\phi_{R}}
\end{align}
where $\mathbf{w}'_{T,S}$ and $\mathbf{w}'_{R}$ are based on the design in (\ref{eq:bf_conv}). For segment $x$, where $x\in\{(T,c),R\}$, $\theta_{x,S}$ is the steering angle for tier-1 beamforming and $\phi_{x}$ is its applied phase for tier-2 beamforming. The compensation gains at $T_{X,c}$ and $R_X$ are given respectively as
\begin{equation}
    \begin{split}
        \alpha_{T,c;m} &= (\mathbf{a}_{T,S}(\theta_{T,c;m}))^T \mathbf{w}'_{T,S}(\theta_{T,c,S})\\
        &=f(\gamma_{T,c,S}-\gamma_{T,c;m}; N_{T,S})/\sqrt{N_{T}},
    \end{split}
    \label{eq:gn1}
\end{equation}
and 
\begin{equation}
    \begin{split}
        \alpha_{R;m} &= (\mathbf{w}'_{R}(\theta_{R,S}))^T\mathbf{a}_{R}(\theta_{R;m})\\
        &=f(\gamma_{R,S}-\gamma_{R;m}; N_{R})/\sqrt{N_{R}}.
    \end{split}
    \label{eq:gn2}
\end{equation}

By following the same design for passive beamforming, the phase shifts introduced by $L_{X,m}$ are given by
\begin{equation}
    \boldsymbol{\Phi}_m = \boldsymbol{\Phi}'_m e^{j\phi_{L,m}}
\end{equation}
where, for a passive steering angle $\theta_{L,m,S}$, $\boldsymbol{\Phi}'_m = diag(\dots,e^{j\gamma_{L,m,S}l}, \dots)_{l\in \mathcal{I}_{N_{L,S}}}$ and $\phi_{L,m}$ is the applied phase for second-tier beamforming. Based on  (\ref{eq:Lx_com_gain}), two responses are given for reflection and the compensation gain of $L_{X,m}$ is given as
\begin{equation}
    \alpha_{m;c} = f(\gamma_{L,m,S}-(\gamma_{L,m;R}+\gamma_{L,m;T,c});N_{L,S}).
\end{equation}
Note that unlike steering angles at $T_{X,c}$ and $R_{X}$, the passive steering angle is not necessarily the same angle towards which the signal is directed.

By ignoring the radiation pattern effect and attenuation due to propagation in (\ref{eq:mpc_gain}), the problem given in (\ref{eq:opt1}) is equivalent to
\begin{equation}
    \begin{split}
        &\boldsymbol{\Omega} = \arg \max_{\boldsymbol{\Omega}} \Big|\sum_{m=1}^{S_L} \sum_{c=1}^{S_T} g_{c,m} e^{jp_{c,m}}\Big|^2\\
        &\hphantom{\boldsymbol{\Omega} }\text{s.t. } |\gamma_{T,c,S}|\leq kq_T, |\gamma_{R,S}|\leq kq_R, |\gamma_{L,m,S}|\leq kq_L,
        \label{eq:opt2}
    \end{split}
\end{equation}
where $\boldsymbol{\Omega} = \left\{ \gamma_{T,c,S},\gamma_{R,S},\gamma_{L,m,S}, \phi_{T,c},\phi_{L,m}\right\}_{m,c}$ is the new parameters vector and
\begin{align}
    g_{c,m} &= \alpha_{T,c;m} \alpha_{R;m} \alpha_{m;c}, \label{eq:mpc_gns}\\
    p_{c,m} &= k(d_{R;m}+d_{T,c;m})+\phi_{T,c}+\phi_{L,m}. \label{eq:mpc_phs}
\end{align}
The phase $\phi_{R}$ is omitted from (\ref{eq:mpc_phs}) as it is independent of the summation indices in (\ref{eq:opt2}). We note that the received signal is a combination of $S_TS_L$ \acp{MPC} (as illustrated in Fig. \ref{fig:segs}), each has magnitude and phase depending on tier-1 and tier-2 beamforming designs, respectively. 

\subsection{Single-Segment Activation}
\label{sec:sngl_seg_act}
Unlike segments at $L_X$ and $R_X$, those at $T_X$ can be switched on and off by means of power allocation. For fixed $S_T$ and $N_{T,s}$, an optimum design leads to no better than maximized gains and completely compensated phase delays for the \acp{MPC} in (\ref{eq:opt2}). Obviously, this is possible when each system unit has one segment (i.e., far-field operation). It is the case of conventional far-field beamforming, where compensation gains are maximized as each beam is focused toward one segment only, and there is one \ac{MPC}, the phase of which can be compensated at any segment.

Changing $S_T$ and $N_{T,S}$ affects $g_{c,m}$ and $p_{c,m}$. As phases have serious impact on the received signal power, the proposed solution next focuses on complete compensation for their effect. Based on (\ref{eq:opt2}) and (\ref{eq:mpc_phs}), we have a total of $S_TS_L$ \acp{MPC}, each with its own phase delay $k(d_{R;m}+d_{T,c;m})$ due to propagation. However, for tier-2 beamforming we have $(S_T+S_L+1)$ degrees of freedom for phase compensation. Therefore, for a complete phase delay compensation, it is required to have $S_T+S_L+1 \geq S_TS_L$. To meet this requirement, a single-segment activation method is proposed where only one segment at $T_X$ is activated by allocating the transmission power only to its elements. Therefore, we have $N_{T,s} = N_T$, and for notational simplicity, the subscript $T,c$ is replaced by $T$.

By changing the number of active elements at $T_X$, we introduce a new optimization variable to the problem in (\ref{eq:opt2}), which is their number $N_{T,S}$. Therefore, the new problem is given as
\begin{equation}
    \begin{split}
        &\{\boldsymbol{\Omega},N_{T,S}\} = \arg \max_{\{\boldsymbol{\Omega},N_{T,S}\}} \Big|\sum_{m=1}^{S_L} g_{m} e^{jp_{m}}\Big|^2\\
        &\hphantom{\{\boldsymbol{\Omega},N_{T,S}\} } \text{s.t. } |\gamma_{T,S}|\leq kq_T, |\gamma_{R,S}|\leq kq_R, |\gamma_{L,m,S}|\leq kq_L,\\
        &\hphantom{\{\boldsymbol{\Omega},N_{T,S}\}  \text{s.t. } } N_{T,S}\leq N_{o},
    \end{split}
    \label{eq:opt3}
\end{equation}
where $N_{o}$ is the maximum number of elements of a $T_X$ segment centered at the antenna center. Given one segment at $T_X$ and $R_X$, we have
\begin{align}
    g_m &= N_{L,S}\alpha_{T;m}\alpha_{R;m},\\
    p_m &= k(d_{T;m}+d_{R;m})+\phi_{L,m}. \label{eq:mpc_pths2}
\end{align}
In comparison with (\ref{eq:mpc_gns}), we note that $\alpha_{m;c}$ is replaced by its maximum value as we choose 
\begin{equation}
    \gamma_{L,m,S} = \gamma_{L,m;T}+\gamma_{L,m;R} \forall m. \label{eq:opt_ans_1}  
\end{equation}
Similar to $\phi_R$ exclusion from (\ref{eq:mpc_phs}), the tier-2 phase at $T_X$ is omitted from (\ref{eq:mpc_pths2}) as it is independent of the summation index in (\ref{eq:opt3}). It can be seen clearly in (\ref{eq:mpc_pths2}) that phase delays are completely compensated by setting tier-2 phases at $L_{X}$ segments such that 
\begin{equation}
\phi_{L,m} = -k(d_{T;m}+d_{R;m})\forall m.
\end{equation}

Based on the design above, tier-1 and tier-2 beamforming at $L_X$ are optimum for the proposed scheme. On the other hand, there is no tier-2 beamforming at $T_X$ and $R_X$, and their compensation gains at $L_X$ segments have some losses depending on their steering angles. As the power captured by $R_X$ is mostly at directions close to its steering angle forming the main lobe of its beam, we neglect compensation gains outside the main lobe by defining the approximation
\begin{equation}
    \alpha'_{R;m} = \begin{cases}
        \alpha_{R;m};& |\gamma_{R,S}-\gamma_{R,m}|\leq 2\pi/N_R\\
        0;& \text{otherwise}.
    \end{cases}
\end{equation}
By adopting the main-lobe approximation given in Appendix \ref{app:n_slo}, we might proceed in solving the beamforming problem by addressing the following problem
\begin{equation}
    \begin{split}
        &\{\boldsymbol{\Omega}',N_{T,S}\} = \arg \max_{\{\boldsymbol{\Omega}',N_{T,S}\}} \sum_{m=1}^{S_L} \alpha'_{T;m}\alpha'_{R;m}\\
        &\hphantom{\{\boldsymbol{\Omega}',N_{T,S}\} } \text{s.t. } |\gamma_{T,S}|\leq kq_T, |\gamma_{R,S}|\leq kq_R,\\
        &\hphantom{\{\boldsymbol{\Omega}',N_{T,S}\}  \text{s.t. } } N_{T,S}\leq N_{o},
        \label{eq:opt4}
    \end{split}
\end{equation}
where $\boldsymbol{\Omega}' = \{ \gamma_{T,S},\gamma_{R,S} \}$ and
\begin{equation}
    \alpha'_{T,m} = f_{\rm approx}(\gamma_{T,s}-\gamma_{T,m};N_{T,S})/\sqrt{N_{T,S}}.
\end{equation}
For a given $N_{T,S}$, steering angles are chosen based on averaging to minimize the compensation lose, so
\begin{align}
    \gamma_{T,S} &= \langle\gamma_{T;m}\rangle_m,\\
    \gamma_{R,S} &= \langle\gamma_{R;m}\rangle_m, \label{eq:opt_ans_end}
\end{align}
where $\langle\cdot\rangle_m$ is the averaging operator over the index $m$. Given $\gamma_{T,S}$, $N_{T,S}$ is found based on the derivative of the objective function of the problem in (\ref{eq:opt4}) such that
\begin{equation}
    \begin{split}
        &\sum_{m=1}^{S_L} \frac{\alpha'_{R;m}}{2\sqrt{N_{T,S}}} \left[ 1 - \frac{5N_{T,S}^2 - 1}{24} \left(\gamma_{T,S} - \gamma_{T;m}\right)^2 \right]\\
        &~~~~~\approx \frac{1}{2\sqrt{N_{T,S}}} \left[ C_0 -\frac{C_1}{5}N_{T,S}^2\right] = 0 \label{eq:der_DK}.
    \end{split}
\end{equation}
Therefore, the value of $N_{T,S}$ that satisfies (\ref{eq:der_DK}) is given as
\begin{equation}
    N_{T,S} = \sqrt{\frac{5C_0}{C_1}},
    \label{eq:opt_nts}
\end{equation}
where 
\begin{align}
    C_0 =& \sum_{m=1}^{S_L}\alpha'_{R;m}\\
    C_1 =& \sum_{m=1}^{S_L}\alpha'_{R;m}\left(\gamma_{T,S} - \gamma_{T;m}\right)^2.
\end{align}

It is important to note that the active segment size depends on the angular span of $L_X$ and beamforming at $R_X$, where the compensation gain of $R_X$ at an $L_X$ segment weights its importance in power reflection. To understand this dependency, consider the following cases with $S_L=N_L$ (i.e., each element at $L_X$ is a segment by itself). By ignoring the operating fields, the first one is for $N_R\rightarrow \infty$, where its compensation gain will be maximum at only one $L_X$ segment and goes to zero for others. In this case, we have $N_{T,S} \rightarrow \infty$, as suggested by (\ref{eq:opt_nts}). It means that the active segment at $T_X$ should have a large number of elements so that all power is focused toward that $L_X$ segment with a maximum $R_X$ compensation gain. Another extreme case is for one-element $L_X$, where we note that (\ref{eq:opt_nts}) always suggests $N_{T,S}\rightarrow \infty$ to concentrate all transmitted power in that element. However, in all cases, we have the upper bound $N_o$ so for $N_{T,S}> N_o$, we have $N_{T,S} = N_o$. For better performance, the value of $N_o$ might be recalculated based on the portion of $L_X$ that is selected by $R_X$ beamforming only; however, this is considered for very large $L_X$ sizes and close distances.

The second case is for $N_R = 1$, which is a worth-investigating practical case. With one antenna element at $R_X$, its compensation gain is equal at all $L_X$ segments and we have
\begin{equation}
    \Big(N_{T,S}\Big)_{\alpha'_{R;m}=1}= \sqrt{\frac{5S_L}{C'_1}},
    \label{eq:opt_nts2}
\end{equation}
where $C'_1 = \sum_{m=1}^{S_L} \left(\gamma_{T,S} - \gamma_{T;m}\right)^2$. A closer look at (\ref{eq:opt_nts2}) tells that $N_{T,S}$ is inversely proportional to the square root of the angular span of $L_X$ segments. The larger the surface (or the deviation of its segments from the center if the coefficient $S_L$ is included), the fewer the antenna elements that should be used. This might be explained by knowing that activating more elements with the beamforming design in (\ref{eq:bf_conv}) gets their signals added constructively at a narrower angular range around the beamforming angle (i.e., smaller beam size). At other directions, the signals with the same power add destructively. Therefore, it makes sense to enlarge the beam size as much as possible so that the signals add constructively in more directions, and this is what (\ref{eq:opt_nts2}) suggests.

\textcolor{hicol}{As seen in (\ref{eq:opt_nts}), the angular span of \ac{IRS} determines the number of elements in the active segment. In this case we considered \acp{ULA}, but for \acp{UPA}, the active segment is two-dimensional. Hence, the extension of the proposed scheme requires finding the number of elements in two dimensions by considering the span of \ac{IRS} over the azimuth and elevation angles in the angular domain. Also, the extension to multiple \acp{IRS} responsible for first-order reflections is straightforward as each \ac{IRS} has its beam at the transmitter. However, for second-order reflections by two \acp{IRS}, the number of \acp{MPC} in the reflected signal increases and exceeds the number of phases that can be applied to them (i.e., the number of design degrees of freedom at the second beamforming tier). Therefore, focusing might not be an optimum solution between two \acp{IRS} and the two tiers need to be considered jointly, which is a case that requires further investigation.}

\textcolor{hicol}{In case of having a non-LoS link between $T_X$ and $L_X$, searching for the best set of scatterers is first required. As mentioned earlier, $L_X$ segments might experience different scatterers with $T_X$, further increasing design complexity. Furthermore, a non-LoS link between $L_X$ and $R_X$ requires an additional search for the best scatterers and increases the reflection order. In such cases, the minimum reflection order is two with additional overhead. On the other hand, first-order reflections are shown in this section to have less overhead and they have much higher power gain, concluding that LoS links with \ac{IRS} are imperative. However, as $L_X$ reflections are controlled and optimized, homogeneous higher-order reflections with no overhead are possible. In this case, the beamforming problem needs further investigation, as discussed earlier.}

\subsection{Numerical Results}
\label{sec:numerical}

\begin{table*}
    \begin{center}
        \begin{tabular}{|c|c c c|c c c|}
            \hline
            \textbf{$L_X$Location:} && $(0,933\lambda)$ &&& $(933\lambda,933\lambda)$ &\\ \hline
        \textbf{$L_X$Size$\backslash$Method}&Span-Based (Prop.) & Far-Field-Based & Main-Lobe-Based &Span-Based (Prop.) & Far-Field-Based & Main-Lobe-Based \\ \hline
        $2048$ & $2$ & $43$ & $4$ & $6$ & $45$ & $9$\\ \hline
        $1024$ & $5$ & $43$ & $8$ & $12$ & $48$ & $20$\\ \hline
        $512$ & $9$ & $43$ & $15$ & $25$ & $50$ & $41$\\ \hline
        $256$ & $18$ & $43$ & $29$ & $51$ & $51$ & $51$\\ \hline
        \end{tabular}
    \end{center}
    \caption{Numbers of $T_X$ active elements under different beamforming schemes for different $L_X$ sizes (in elements) and positions.}
    \label{tbl:tbl}
\end{table*}

As detailed previously, $L_X$ was shown to be part of the channel as a scattering cluster with its own \ac{MPC} as a specular component of the received signal by $R_X$. In this section, the performance of this cluster is investigated by means of simulations for the proposed beamforming scheme. 

Cascaded beamforming was shown to have two cascaded parts: power collection and reflection. First, power collection is investigated, where $L_X$ acts as a receiver with analog beamforming and it is desired to deliver it as much power as possible. Next, power reflection is addressed, where the achieved throughput by $L_X$ specular component is shown for different scenarios. To focus on the performance of the controlled scattering cluster, we assume no scatterers in the environment except the controlled ones. Propagation parameters are $a_{\rm att} = 2$ and $b_{\rm att} = 2k$ for LoS links between system units \cite{basar2020indoor}. The power radiation pattern of $L_X$ elements is given as
\begin{equation}
    B_e(\theta) = 2(2q+1) \cos^{2q}(\theta)
\end{equation}
where $\theta$ is the element broadside angle and $q\approx0.285$ is an introduced parameter to ensure power conservation \cite{ellingson2019path}. Distances and sizes are given in terms of the wavelength, but the operating frequency is $28$GHz. The transmitted power is set to $0$dBm, and the noise floor is $-90$dBm. Antenna element spacing is selected for all system units to be $\lambda/2$. Simulations are conducted in the 2D space. Unless otherwise specified, $T_X$ has $64$ elements and it is laid on the x-axis with its center at $(0,0)$. All system units are placed horizontally if orientation is not mentioned.

\begin{figure}[t]
    \centering
\includegraphics[scale=0.3]{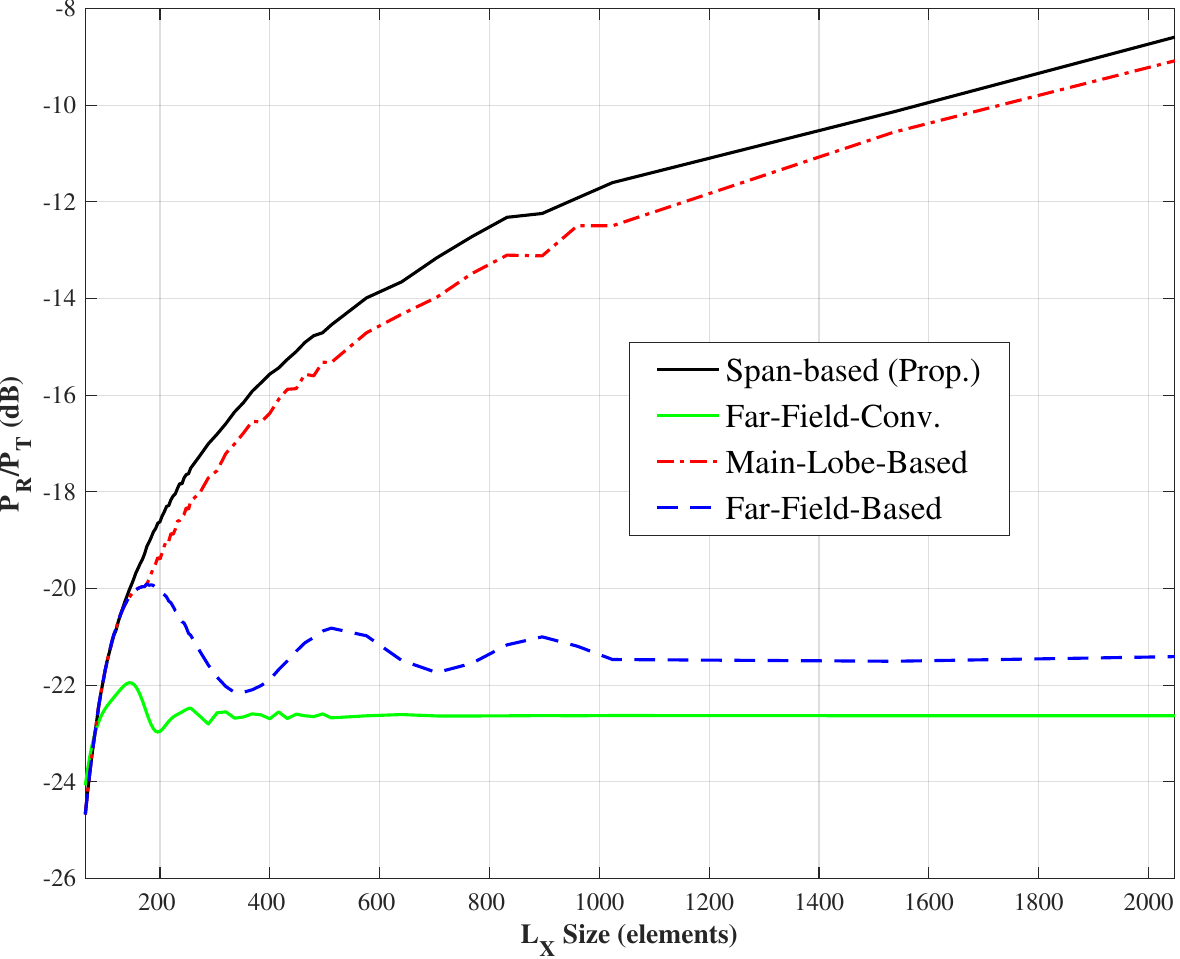}
\caption{Path Gain for different beamforming schemes with $L_X$ located perpendicular to $T_X$ at $(0,933\lambda)$.}
\label{fig:sims3a}
\end{figure}

\begin{figure}[t]
    \centering
\includegraphics[scale=0.3]{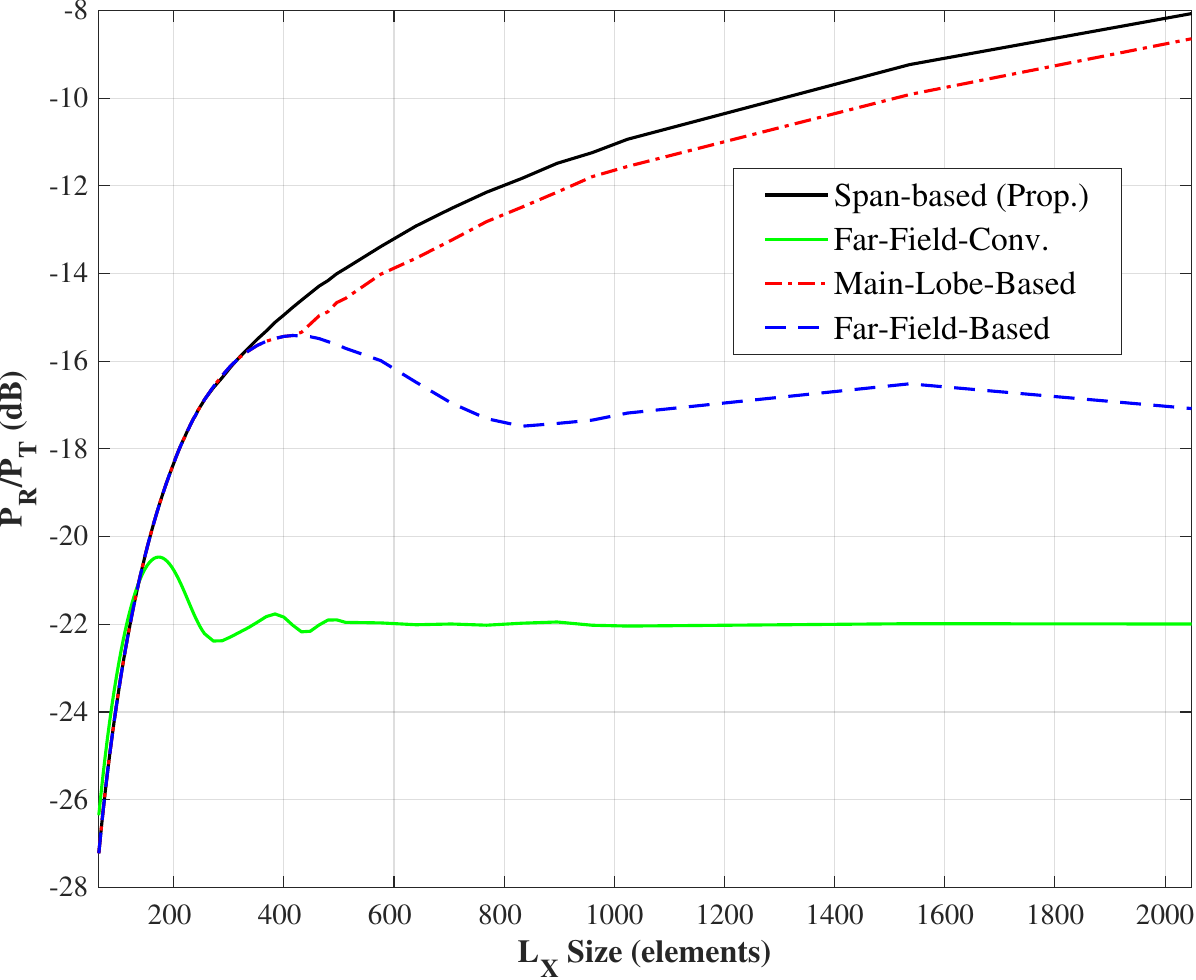}
\caption{Path gain for different beamforming schemes with $L_X$ located at $(933\lambda,933\lambda)$.}
\label{fig:sims3b}
\end{figure}

For one-element $R_X$, Fig. \ref{fig:sims3a} and Fig. \ref{fig:sims3b} show the path gain between $T_X$ and $L_X$ for different beamforming schemes under different $L_X$ sizes and positioning. In Fig. \ref{fig:sims3a}, $L_X$ is located at $(0,933\lambda)$m, or $(0,10)$m at the given operating frequency. For the plots in Fig. \ref{fig:sims3b}, on the other hand, $L_X$ is located at $(933\lambda,933\lambda)$m. Four cases are considered as follows. The one denoted by "Span-Based" is for the proposed scheme in the previous section, and "Far-Field-Conv" is given for the conventional beamforming that assumes far-field operation for all system units. The "Far-Field-Based" case is shown for a fixed active segment size depending on the distances with $L_X$ elements regardless of its size. Finally, "Main-Lobe-Based" is given for the segment size selection that ensures covering $L_X$ only by the main lobe, as long as it does not exceed the far-field-restricted segment size. 

For IRS-size-independent schemes (i.e., far-field-based and far-field-conv methods), we note that the received signal power converges to a constant level. As $L_X$ becomes larger, it will capture more power, regardless of how signals are precoded and combined. This continues until it becomes large enough to capture all main lobes of the kernels given in (\ref{eq:segs_tx_2}). For the main-Lobe-Based case, it is only one beam. The convergence power level depends on $T_X$ and $L_X$ angular spans as viewed by each other and the number of active elements at $T_X$. When $T_X$ has a smaller angular span seen by $L_X$, $L_X$ will be closer to operate in its far field. In the far field, conventional beamforming at $T_X$ is optimum as the obtained power level is the best that can be captured.

For IRS-size-dependent active segmentation (i.e., span-based and main-lobe-based methods), we note that the power level keeps increasing. The reason is that these methods reduce the number of active elements as $L_X$ gets larger, giving a chance to increase the main lobe size as discussed earlier. The main-lobe-based method shows how the proposed method performs just better than heuristically choosing to cover $L_X$ surface by the main lobe of the active segment. It is only the case for one-element $R_X$; otherwise, the main-lobe-based method fails to account for $R_X$ beamforming. Note here that using fewer elements serves better for large $L_X$ sizes. Table \ref{tbl:tbl} shows how the number of elements decreases for wider angular span by $L_X$. For example, based on the proposed method, $9$ elements are recommended to deliver more power to an $L_X$ with $512$ elements located at $(0,933\lambda)$m, though $43$ elements are possible. Figures \ref{fig:sims3a} and \ref{fig:sims3b} show the difference between these two schemes, which is significant for larger $L_X$ sizes. For smaller $L_X$ spans, the two methods converge to the far-field-based method. At $(933\lambda,933\lambda)$m, $L_X$ has a smaller angular span and that explains why more elements are involved in both span-based and main-lobe-based methods.

\begin{figure}[t]
    \centering
\includegraphics[scale=0.3]{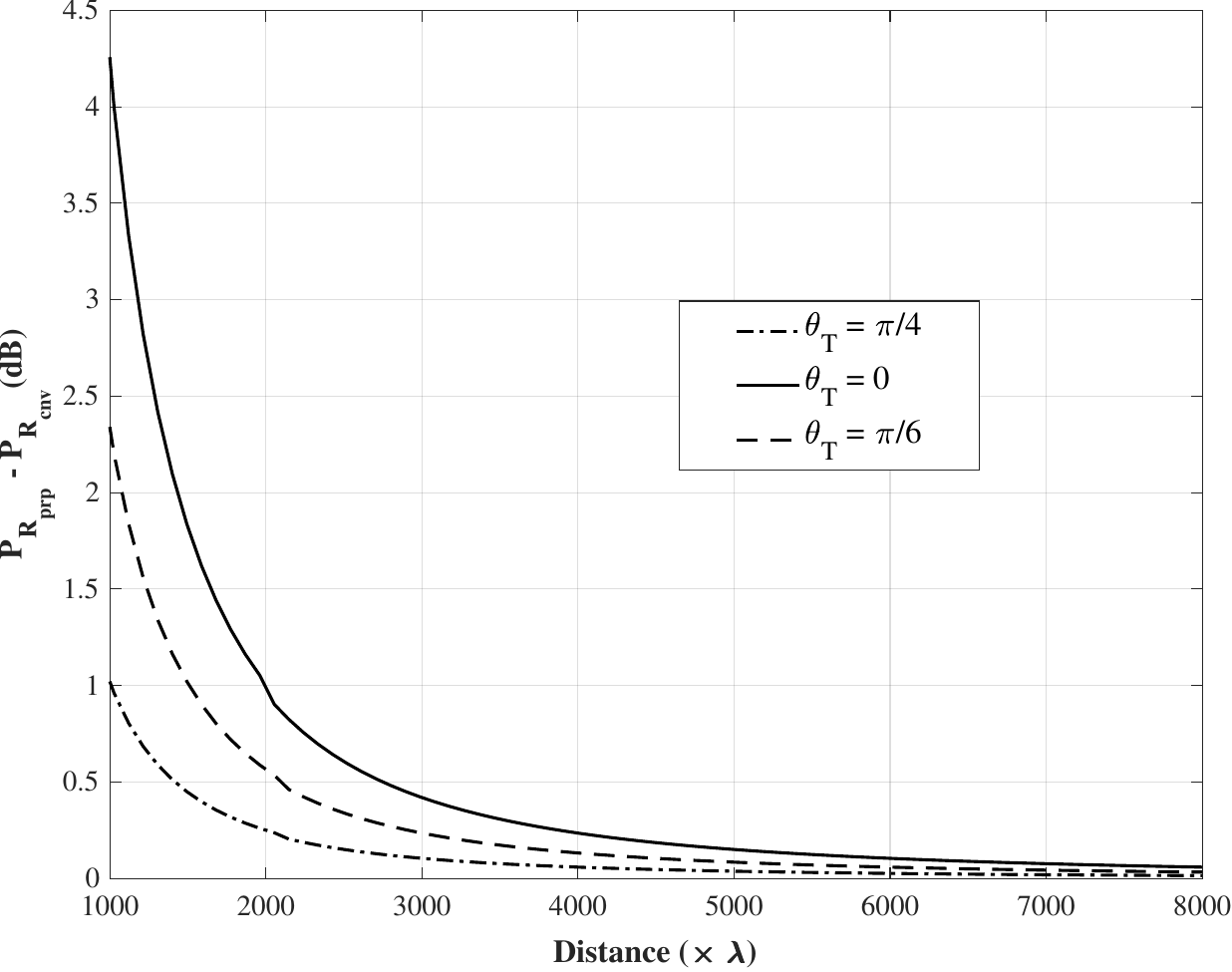}
\caption{Convergence of the proposed beamforming scheme to the conventional far-field scheme with distance. }
\label{fig:sims2a}
\end{figure}

The convergence of the proposed beamforming scheme to the conventional far-field scheme with distance is shown in Fig. \ref{fig:sims2a}. For demonstration purposes, $T_X$ has only one element but $R_X$ has $128$ elements. The difference in power delivered by the two schemes decreases as $R_X$ moves away from $T_X$ since it becomes one segment and the proposed scheme decreases to the conventional one. However, for larger $\theta_T$, which is the angular location of $R_X$ with respect to the positive y-axis, the convergence is met faster. The reason is that $R_X$ will have a smaller angular span seen by $T_X$, and as a result, its far-field boundary distance with $T_X$ is smaller.

\begin{figure}[t]
    \centering
\includegraphics[scale=0.3]{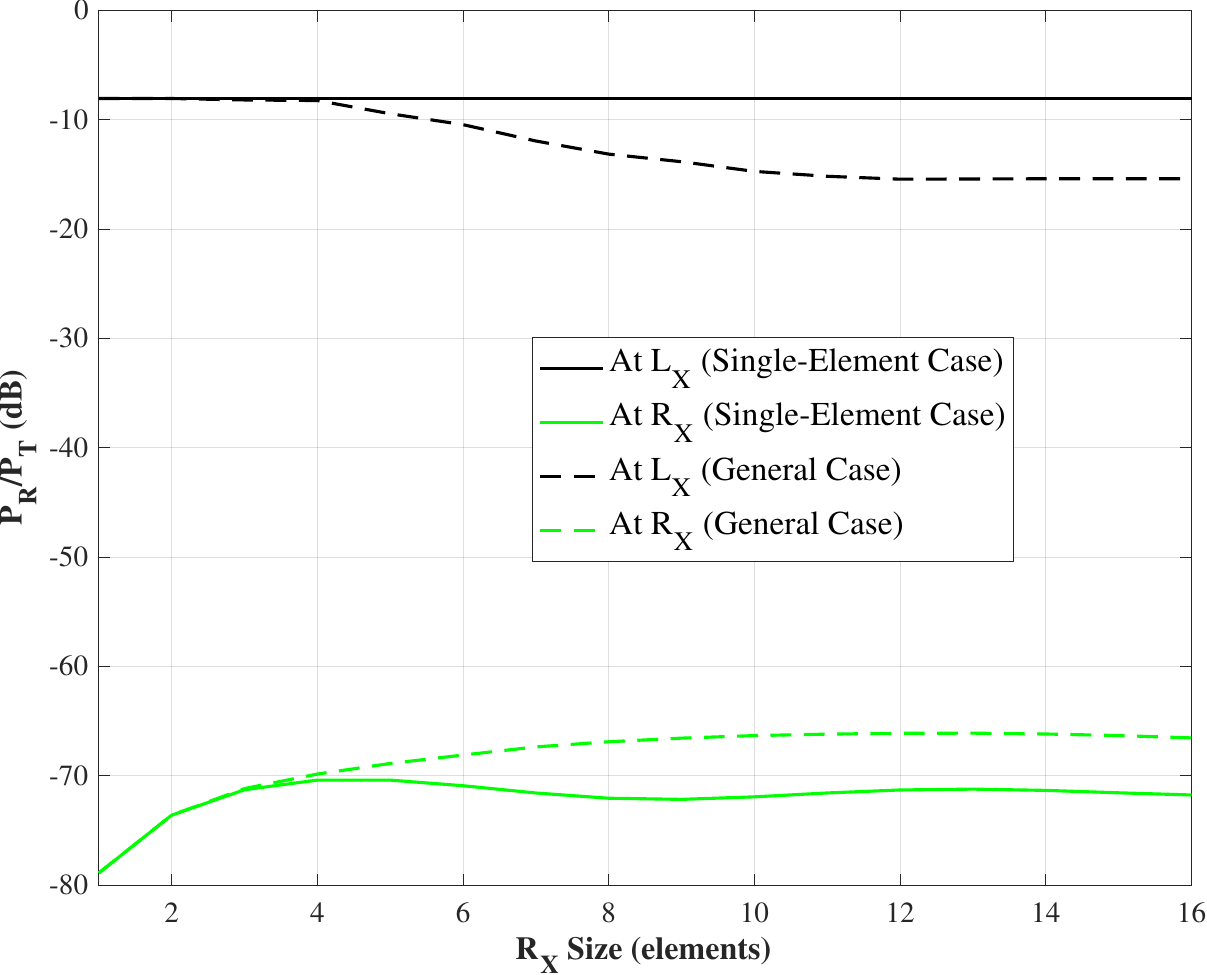}
\caption{Power Levels at $L_X$ and $R_X$ with and without $R_X$ beamforming effect.}
\label{fig:sims2b}
\end{figure}

The number of active elements at $T_X$ and $L_X$ size were shown to be inversely proportional in the case of having one-element $R_X$. However, this is not the case when we have multiple elements at $R_X$ as its beamforming has a selection effect over $L_X$ segments. Figure \ref{fig:sims2b} shows this effect. Both $L_X$ and $R_X$ are fixed in location at $(467\lambda,467\lambda)$ and $(467\lambda,0)$, respectively, and $L_X$ size is $512\lambda$. As the number of $R_X$ elements increases, its compensation gain at those $L_X$ segments far from its center is low. If the active segment at $T_X$ is generated based on (\ref{eq:opt_nts2}) as shown for the "Single-Element" case, the power received by $R_X$ will be less compared with generation based on (\ref{eq:opt_nts}), which is the "General" case. The reason is that power is spread over the whole $L_X$ surface in the signal-element case, though only a portion of the surface contributes to power reflection. On the other hand, when $R_X$ beamforming is considered, despite having less power collected by $L_X$, more power is received at $R_X$. This is not to be confused with the desire to have the main lobe of the beam by $T_X$ as large as possible.

\begin{figure}[t]
    \centering
\includegraphics[scale=0.3]{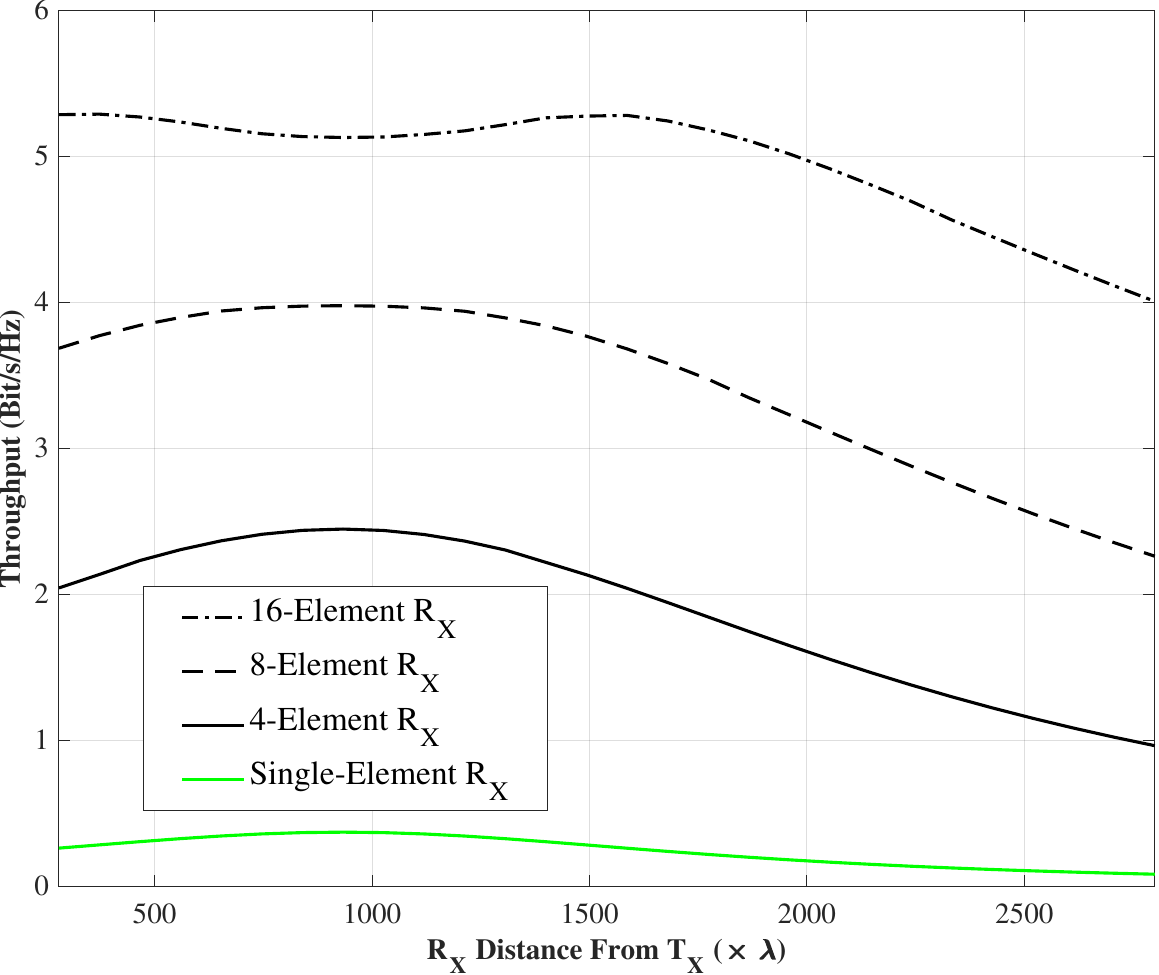}
\caption{\color{hicol}Throughput achieved by the \ac{MPC} reflected by $L_X$ for different sizes of $R_X$ at different locations.}
\label{fig:simsa}
\end{figure}

The throughput achieved by the \ac{MPC} of $L_X$ at $R_X$ with different sizes is shown in Fig. \ref{fig:simsa}. $L_X$ is located at $(933\lambda,933\lambda)$ and $R_X$ is moving from $(280\lambda,0)$ to $(2800\lambda,0)$. As might be expected, when getting closer to $L_X$, the throughput increases. This is clearly the case for small number of elements at $R_X$; however, as their number increases, due to receive beamforming, maximum throughput is achieved at other locations. At a perpendicular location, $L_X$ will have the maximum angular span at $R_X$, and with more elements at $L_X$ or $R_X$, there are more chances to get some $L_X$ segments canceled by $R_X$ beamforming. As $R_X$ moves away, however, $L_X$ occupies smaller angular span and more of its segments get accepted by beamforming at $R_X$.

\begin{figure}[t]
    \centering
\includegraphics[scale=0.3]{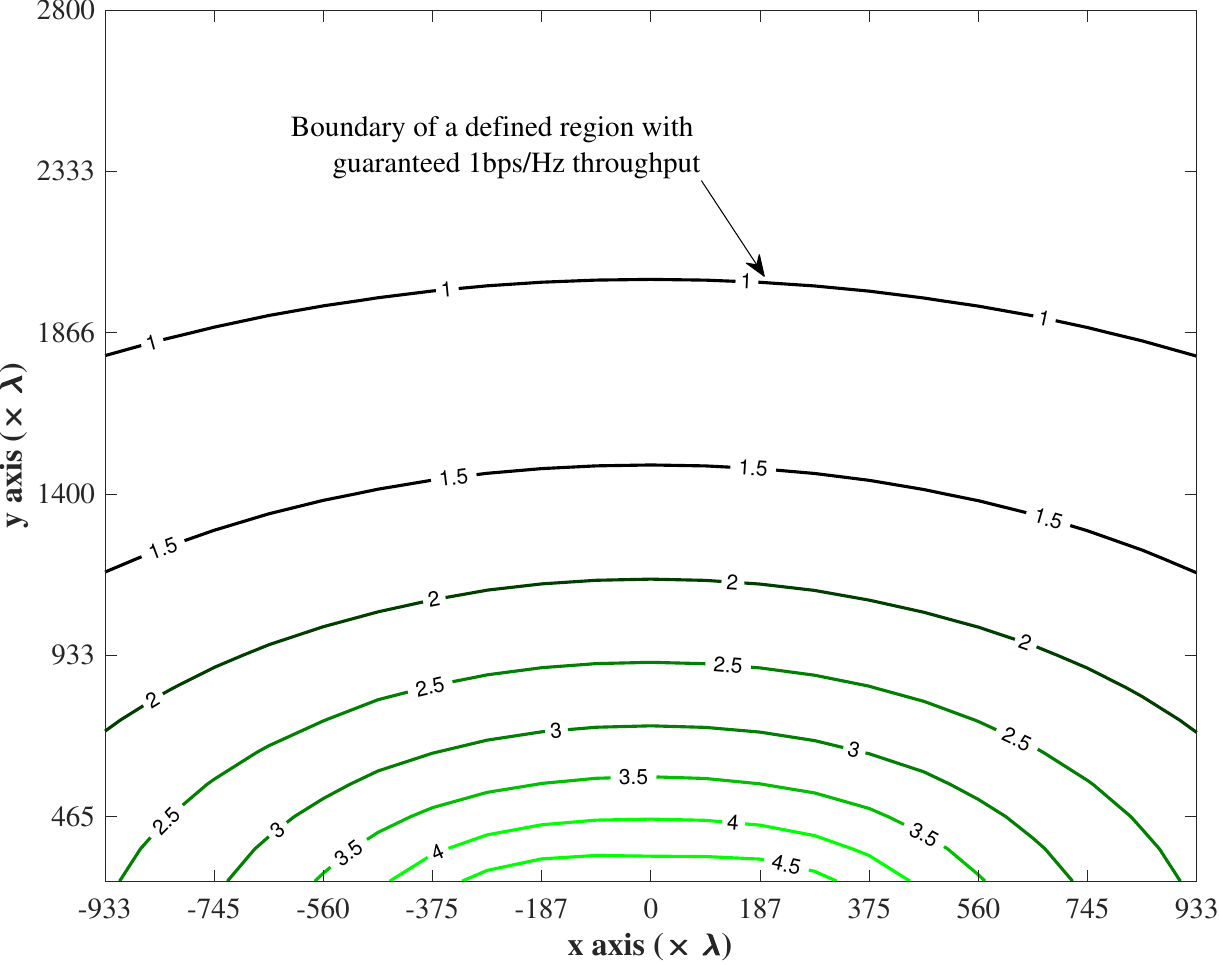}
\caption{\color{hicol} Boundaries of defined regions with different levels of guaranteed throughputs.}
\label{fig:simsb}
\end{figure}

Defined regions might be covered by one or more quasi-LoS links through \acp{IRS}. Figure \ref{fig:simsb} shows an example of the coverage provided by a single $L_X$ within a given region. For this specific example, $L_X$ size is $256\lambda$ and it is centered at $(0,0)$. $T_X$ is located at $(-933\lambda,933\lambda)$. The throughput is calculated for a four-element $R_X$. A contour line represents the boundary of a region within which the capacity is guaranteed to be above its label number. We note that closer to $L_X$, the boundaries are not circular for the same reason explained earlier. However, far from $L_X$, this deviation is less, and the boundaries become more circular. Finally, the capacity drop with distance is not linear at all angles.

\section{Conclusion}
\label{sec:cnc}

In this work, \ac{IRS} was investigated as part of the channel. Similar to any object in the environment, it acts as a scattering cluster but a controlled one. We have shown the way this control is possible and the nature of its reflected signal. By segmentation, each group of \ac{IRS} elements is indeed a scatterer in the cluster. 

\color{hicol}
A new model was proposed for the channel as a whole based on the classification of the \ac{MPC} paths. By segmentation, we showed that the beam in the near field is a combination of kernels. Based on this model, a beamforming scheme was proposed to add those kernels constructively and maximize the received power. 

The developed beam-space model shows how we really have a degree of control over the wireless channel, and how the channel can be converted from a problem to be part of the design. With its directed \acp{MPC}, the controlled cluster could open the door for new solutions in addressing both classical and emerging problems for future wireless network generations.

\color{black}

\appendices
\section{Main-Lobe Approximation}
\label{app:n_slo}
Given that $e^{jx} = \sum_{s=0}^{\infty}(jx)^s/s!$, we have
\begin{equation}
    \begin{split}
        f(\theta;N) &= \sum_{l\in \mathcal{I}_N} \sum_{s=0}^{\infty}\frac{(j\theta l)^s}{s!} \\
        &= \sum_{l\in \mathcal{I}_N} \sum_{k=0}^{\infty} \frac{(j\theta l)^{2k}}{(2k)!}
    \end{split}
    \label{eq:approx}
\end{equation}
The second equality says that the imaginary part of the series cancels out due to the interval $\mathcal{I}_{N}$ symmetry around the zero. The main lobe of $f(\theta;N)$ carries most of its power. For our purpose in solving the beamforming problem, it is approximated by taking only $k\leq1$ in (\ref{eq:approx}), so we have the second-order polynomial 
\begin{equation}
    \begin{split}
        f_{\rm approx }(\theta;N)&= N - \theta^2\sum_{l\in \mathcal{I}_N} \frac{l^2}{2}\\
        &= {N - \frac{N^3-N}{24}\theta^2},
        \label{eq:dk_approx}
    \end{split}
\end{equation}
where in the second line we used the equality $\sum_{s=1}^{M}s^2 = M(M+1)(2M+1)/6$.

\bibliographystyle{IEEEtran}
\bibliography{refs}   

\begin{thebibliography}{10}
\providecommand{\url}[1]{#1}
\csname url@samestyle\endcsname
\providecommand{\newblock}{\relax}
\providecommand{\bibinfo}[2]{#2}
\providecommand{\BIBentrySTDinterwordspacing}{\spaceskip=0pt\relax}
\providecommand{\BIBentryALTinterwordstretchfactor}{4}
\providecommand{\BIBentryALTinterwordspacing}{\spaceskip=\fontdimen2\font plus
\BIBentryALTinterwordstretchfactor\fontdimen3\font minus
  \fontdimen4\font\relax}
\providecommand{\BIBforeignlanguage}[2]{{%
\expandafter\ifx\csname l@#1\endcsname\relax
\typeout{** WARNING: IEEEtran.bst: No hyphenation pattern has been}%
\typeout{** loaded for the language `#1'. Using the pattern for}%
\typeout{** the default language instead.}%
\else
\language=\csname l@#1\endcsname
\fi
#2}}
\providecommand{\BIBdecl}{\relax}
\BIBdecl

\bibitem{8466374}
C.~{Liaskos}, S.~{Nie} \emph{et~al.}, ``A new wireless communication paradigm
  through software-controlled metasurfaces,'' \emph{IEEE Commun. Mag.},
  vol.~56, no.~9, pp. 162--169, 2018.

\bibitem{8796365}
E.~{Basar}, M.~{Di Renzo} \emph{et~al.}, ``Wireless communications through
  reconfigurable intelligent surfaces,'' \emph{IEEE Access}, vol.~7, pp.
  116\,753--116\,773, 2019.

\bibitem{Renzo_2020}
M.~{Di Renzo}, A.~{Zappone} \emph{et~al.}, ``Smart radio environments empowered
  by reconfigurable intelligent surfaces: How it works, state of research, and
  the road ahead,'' \emph{IEEE J. Sel. Areas Commun.}, vol.~38, no.~11, pp.
  2450--2525, 2020.

\bibitem{zhang2020capacity}
S.~Zhang and R.~Zhang, ``Capacity characterization for intelligent reflecting
  surface aided {MIMO} communication,'' \emph{IEEE J. Sel. Areas Commun.},
  vol.~38, no.~8, pp. 1823--1838, 2020.

\bibitem{qiao2020secure}
J.~Qiao and M.-S. Alouini, ``Secure transmission for intelligent reflecting
  surface-assisted {mmWave} and terahertz systems,'' \emph{IEEE Wireless
  Commun. Lett.}, vol.~9, no.~10, pp. 1743--1747, 2020.

\bibitem{shen2019secrecy}
H.~Shen, W.~Xu \emph{et~al.}, ``Secrecy rate maximization for intelligent
  reflecting surface assisted multi-antenna communications,'' \emph{IEEE
  Commun. Lett.}, vol.~23, no.~9, pp. 1488--1492, 2019.

\bibitem{cui2019secure}
M.~Cui, G.~Zhang, and R.~Zhang, ``Secure wireless communication via intelligent
  reflecting surface,'' \emph{IEEE Wireless Commun. Lett.}, vol.~8, no.~5, pp.
  1410--1414, 2019.

\bibitem{8811733}
Q.~{Wu} and R.~{Zhang}, ``Intelligent reflecting surface enhanced wireless
  network via joint active and passive beamforming,'' \emph{IEEE Trans.
  Wireless Commun.}, vol.~18, no.~11, pp. 5394--5409, 2019.

\bibitem{8741198}
C.~{Huang}, A.~{Zappone} \emph{et~al.}, ``Reconfigurable intelligent surfaces
  for energy efficiency in wireless communication,'' \emph{IEEE Trans. Wireless
  Commun.}, vol.~18, no.~8, pp. 4157--4170, 2019.

\bibitem{wang2020intelligent}
P.~Wang, J.~Fang \emph{et~al.}, ``Intelligent reflecting surface-assisted
  millimeter wave communications: Joint active and passive precoding design,''
  \emph{IEEE Trans. Veh. Technol.}, 2020.

\bibitem{kammoun2020asymptotic}
A.~Kammoun, A.~Chaaban \emph{et~al.}, ``Asymptotic max-min {SINR} analysis of
  reconfigurable intelligent surface assisted miso systems,'' \emph{IEEE Trans.
  Wireless Commun.}, vol.~19, no.~12, pp. 7748--7764, 2020.

\bibitem{ding2020simple}
Z.~Ding and H.~V. Poor, ``A simple design of {IRS}-{NOMA} transmission,''
  \emph{IEEE Commun. Lett.}, vol.~24, no.~5, pp. 1119--1123, 2020.

\bibitem{hou2019reconfigurable}
T.~Hou, Y.~Liu \emph{et~al.}, ``Reconfigurable intelligent surface aided {NOMA}
  networks,'' \emph{IEEE J. Sel. Areas Commun.}, vol.~38, no.~11, pp.
  2575--2588, 2020.

\bibitem{9198898}
A.~{Almohamad}, A.~M. {Tahir} \emph{et~al.}, ``Smart and secure wireless
  communications via reflecting intelligent surfaces: A short survey,''
  \emph{IEEE Open J. Commun. Soc.}, vol.~1, pp. 1442--1456, 2020.

\bibitem{9122596}
S.~Gong, X.~Lu \emph{et~al.}, ``Toward smart wireless communications via
  intelligent reflecting surfaces: A contemporary survey,'' \emph{IEEE Commun.
  Surveys Tuts.}, vol.~22, no.~4, pp. 2283--2314, 2020.

\bibitem{9133142}
C.~{You}, B.~{Zheng}, and R.~{Zhang}, ``Channel estimation and passive
  beamforming for intelligent reflecting surface: Discrete phase shift and
  progressive refinement,'' \emph{IEEE J. Sel. Areas Commun.}, vol.~38, no.~11,
  pp. 2604--2620, 2020.

\bibitem{zegrar2020general}
S.~E. Zegrar, L.~Afeef, and H.~Arslan, ``A general framework for {RIS}-aided
  {mmWave} communication networks: Channel estimation and mobile user
  tracking,'' \emph{arXiv preprint arXiv:2009.01180}, 2020.

\bibitem{zegrar2020reconfigurable}
------, ``Reconfigurable intelligent surfaces ({RIS}): Channel model and
  estimation,'' \emph{arXiv preprint arXiv:2010.05623}, 2020.

\bibitem{9127834}
S.~{Liu}, Z.~{Gao} \emph{et~al.}, ``Deep denoising neural network assisted
  compressive channel estimation for {mmWave} intelligent reflecting
  surfaces,'' \emph{IEEE Trans. Veh. Technol.}, vol.~69, no.~8, pp. 9223--9228,
  2020.

\bibitem{8879620}
Z.~{He} and X.~{Yuan}, ``Cascaded channel estimation for large intelligent
  metasurface assisted massive {MIMO},'' \emph{IEEE Wireless Commun. Lett.},
  vol.~9, no.~2, pp. 210--214, 2020.

\bibitem{8424015}
C.-X. Wang, J.~Bian \emph{et~al.}, ``A survey of {5G} channel measurements and
  models,'' \emph{IEEE Commun. Surveys Tuts.}, vol.~20, no.~4, pp. 3142--3168,
  2018.

\bibitem{6736750}
W.~Roh, J.-Y. Seol \emph{et~al.}, ``Millimeter-wave beamforming as an enabling
  technology for {5G} cellular communications: theoretical feasibility and
  prototype results,'' \emph{IEEE Commun. Mag.}, vol.~52, no.~2, pp. 106--113,
  2014.

\bibitem{7109864}
T.~S. {Rappaport}, G.~R. {MacCartney} \emph{et~al.}, ``Wideband millimeter-wave
  propagation measurements and channel models for future wireless communication
  system design,'' \emph{IEEE Trans. Commun.}, vol.~63, no.~9, pp. 3029--3056,
  2015.

\bibitem{peter2017measurement}
M.~Peter, K.~Haneda \emph{et~al.}, ``Measurement results and final {mmMAGIC}
  channel models,'' \emph{Deliverable D2}, vol.~2, 2017.

\bibitem{maltsev2009experimental}
A.~Maltsev, R.~Maslennikov \emph{et~al.}, ``Experimental investigations of
  60{GHz} {WLAN} systems in office environment,'' \emph{IEEE J. Sel. Areas
  Commun.}, vol.~27, no.~8, pp. 1488--1499, 2009.

\bibitem{zhang2020prospective}
J.~Zhang, E.~Björnson \emph{et~al.}, ``Prospective multiple antenna
  technologies for beyond {5G},'' \emph{IEEE J. Sel. Areas Commun.}, vol.~38,
  no.~8, pp. 1637--1660, 2020.

\bibitem{1033686}
A.~M. {Sayeed}, ``Deconstructing multiantenna fading channels,'' \emph{IEEE
  Trans. Signal Process.}, vol.~50, no.~10, pp. 2563--2579, 2002.

\bibitem{tse2005fundamentals}
D.~Tse and P.~Viswanath, \emph{Fundamentals of Wireless Communication}, ser.
  Wiley series in telecommunications.\hskip 1em plus 0.5em minus 0.4em\relax
  Cambridge University Press, 2005.

\bibitem{8207426}
I.~A. {Hemadeh}, K.~{Satyanarayana} \emph{et~al.}, ``Millimeter-wave
  communications: Physical channel models, design considerations, antenna
  constructions, and link-budget,'' \emph{IEEE Commun. Surveys Tuts.}, vol.~20,
  no.~2, pp. 870--913, 2018.

\bibitem{basar2020indoor}
E.~Basar, I.~Yildirim, and F.~Kilinc, ``Indoor and outdoor physical channel
  modeling and efficient positioning for reconfigurable intelligent surfaces in
  {mmWave} bands,'' \emph{IEEE Trans. Commun.}, 2021.

\bibitem{wang2020joint}
W.~Wang and W.~Zhang, ``Joint beam training and positioning for intelligent
  reflecting surfaces assisted millimeter wave communications,'' \emph{IEEE
  Trans. Wireless Commun.}, vol.~20, no.~10, pp. 6282--6297, 2021.

\bibitem{ellingson2019path}
S.~W. Ellingson, ``Path loss in reconfigurable intelligent surface-enabled
  channels,'' \emph{arXiv preprint arXiv:1912.06759}, 2019.

\bibitem{tang2020wireless}
W.~Tang, M.~Z. Chen \emph{et~al.}, ``Wireless communications with
  reconfigurable intelligent surface: Path loss modeling and experimental
  measurement,'' \emph{IEEE Trans. Wireless Commun.}, 2020.

\bibitem{8936989}
{\"O}.~{Özdogan}, E.~{Björnson}, and E.~G. {Larsson}, ``{Intelligent
  Reflecting Surfaces: Physics, Propagation, and Pathloss Modeling},''
  \emph{IEEE Wireless Commun. Lett.}, vol.~9, no.~5, pp. 581--585, 2020.

\bibitem{gradoni2021end}
G.~Gradoni and M.~Di~Renzo, ``End-to-end mutual coupling aware communication
  model for reconfigurable intelligent surfaces: An electromagnetic-compliant
  approach based on mutual impedances,'' \emph{IEEE Wireless Commun. Lett.},
  2021.

\bibitem{yang2019surface}
F.~Yang and Y.~Rahmat-Samii, \emph{Surface electromagnetics: with applications
  in antenna, microwave, and optical engineering}.\hskip 1em plus 0.5em minus
  0.4em\relax Cambridge University Press, 2019.

\bibitem{8736783}
B.~{Friedlander}, ``Localization of signals in the near-field of an antenna
  array,'' \emph{IEEE Trans. Signal Process.}, vol.~67, no.~15, pp. 3885--3893,
  2019.

\bibitem{7942128}
K.~T. {Selvan} and R.~{Janaswamy}, ``Fraunhofer and fresnel distances: Unified
  derivation for aperture antennas.'' \emph{IEEE Antennas Propag. Mag.},
  vol.~59, no.~4, pp. 12--15, 2017.

\bibitem{6717211}
O.~E. {Ayach}, S.~{Rajagopal} \emph{et~al.}, ``Spatially sparse precoding in
  millimeter wave {MIMO} systems,'' \emph{IEEE Trans. Wireless Commun.},
  vol.~13, no.~3, pp. 1499--1513, 2014.

\bibitem{6484896}
J.~{Brady}, N.~{Behdad}, and A.~M. {Sayeed}, ``Beamspace {MIMO} for
  millimeter-wave communications: System architecture, modeling, analysis, and
  measurements,'' \emph{IEEE Trans. Antennas Propag.}, vol.~61, no.~7, pp.
  3814--3827, 2013.

\bibitem{7448873}
A.~Alkhateeb and R.~W. Heath, ``Frequency selective hybrid precoding for
  limited feedback millimeter wave systems,'' \emph{IEEE Trans. Commun.},
  vol.~64, no.~5, pp. 1801--1818, 2016.

\bibitem{maltsev2017channel}
A.~Maltsev \emph{et~al.}, ``Channel models for {IEEE} 802.11ay,'' \emph{IEEE
  document 802.11-15/1150r9}, 2017.

\bibitem{6848765}
A.~{Meijerink} and A.~F. {Molisch}, ``On the physical interpretation of the
  saleh–valenzuela model and the definition of its power delay profiles,''
  \emph{IEEE Trans. Antennas Propag.}, vol.~62, no.~9, pp. 4780--4793, 2014.

\bibitem{gradshteyn2014table}
I.~S. Gradshteyn and I.~M. Ryzhik, \emph{Table of integrals, series, and
  products}.\hskip 1em plus 0.5em minus 0.4em\relax Academic press, 2014.

\bibitem{6810277}
X.~Gao, F.~Tufvesson, and O.~Edfors, ``Massive {MIMO} channels — measurements
  and models,'' in \emph{{2013 Asilomar Conference on Signals, Systems and
  Computers}}, 2013, pp. 280--284.

\bibitem{8290972}
C.~F. López and C.-X. Wang, ``Novel 3-{D} non-stationary wideband models for
  massive {MIMO} channels,'' \emph{IEEE Trans. Wireless Commun.}, vol.~17,
  no.~5, pp. 2893--2905, 2018.

\end{thebibliography}

\end{document}